\documentclass[a4paper,11pt]{article}

\pdfoutput=1

\usepackage[T1]{fontenc}
\usepackage[utf8]{inputenc}
\usepackage[english]{babel}

\usepackage{amsmath,amssymb,amsfonts,mathtools}
\usepackage{amsthm}
\usepackage{latexsym}
\usepackage[mathcal]{euscript}
\usepackage{mathrsfs}
\usepackage{bm}

\usepackage{graphicx}
\usepackage{epsfig}
\usepackage{subfig}
\usepackage{psfrag}
\usepackage{array}
\usepackage{float}
\usepackage{cancel}
\usepackage{slashed}
\usepackage{appendix}

\usepackage[normalem]{ulem}
\usepackage{xcolor}
\usepackage{microtype}
\usepackage{enumitem}

\usepackage{jheppub}

\usepackage{hyperref}
\usepackage{cleveref}

\hypersetup{
    colorlinks=true,
    linkcolor=blue!55!black,
    citecolor=blue!55!black,
    urlcolor=blue!55!black
}

\numberwithin{equation}{section}

\newcommand{\dd}{\mathrm{d}}

\newcommand{\cE}{\mathcal{E}}
\newcommand{\cN}{\mathcal{N}}

\newcommand{\cM}{\mathcal{M}}

\newcommand{\be}{\begin{equation}}
\newcommand{\ee}{\end{equation}}

\newcommand{\bea}{\begin{eqnarray}}
\newcommand{\eea}{\end{eqnarray}}

\newcommand{\bi}{\begin{itemize}}
\newcommand{\ei}{\end{itemize}}

\begin{document}

\title{Holographic Drag Force and the Entanglement Membrane: Geometric Reconstruction and Information Spreading}

\abstract{A central question in holography is how different boundary observables encode and reconstruct the geometry of the gravitational dual. In this work, we uncover an exact geometric correspondence between two seemingly distinct aspects of strongly coupled dynamics: the dissipative drag force experienced by a moving heavy particle(quark) and the entanglement membrane governing the propagation of quantum information. We show that both constructions retain the same underlying  structure of the geometry, leading to an invertible transformation between the finite-velocity drag force and the geometric membrane tension. This correspondence reveals that nonlinear momentum loss encodes detailed information about entanglement dynamics: the leading nonlinear correction determines the butterfly velocity, while higher-order terms successively probe the local geometry of the membrane near the horizon. We further connect this hierarchy to scalar pole-skipping, establishing an all-order relation between nonlinear dissipation and near-horizon spectral data. The shared geometric structure also has implications for quantum information recovery, with the quintic drag coefficient determining the leading dependence of information velocity on initial entanglement. Conversely, the complete information-velocity curve provides a reconstruction of the reduced bulk geometry along the physical interior membrane branch, complementing the exterior reconstruction from drag. We show that stochastic momentum broadening supplies the additional radial information needed to reconstruct the complete exterior metric up to an overall scale. Our results thus reveal how dissipation, entanglement dynamics, and spectral properties provide complementary access to holographic geometry, establishing a unified framework for understanding the geometric information encoded in distinct boundary observables.}

\author[a,b]{Dmitry S. Ageev}

\affiliation[a]{Department of Mathematical Methods for Quantum Technologies, Steklov Mathematical Institute of Russian Academy of Sciences, Gubkina str. 8, Moscow 119991, Russia}
\affiliation[b]{Institute for Theoretical and Mathematical Physics, Lomonosov Moscow State University, 119991 Moscow, Russia}

\emailAdd{ageev@mi-ras.ru}

\maketitle

\section{Introduction}
\label{sec:introduction}

Effective field theory provides a systematic way of identifying the structures that survive coarse graining, allowing long-distance dynamics to be described without resolving every microscopic degree of freedom. In nonequilibrium systems, this perspective is particularly valuable: hydrodynamics organizes the evolution of conserved densities, while effective descriptions of dissipation and fluctuations identify the response data that remain relevant at late times \cite{CrossleyGloriosoLiu2017}. The dynamics of entanglement raises a related but distinct question. Quantum information is not a locally conserved density, yet its redistribution in chaotic many-body systems can exhibit simple large-scale behavior. Membrane theories of entanglement dynamics provide an effective description of this behavior in terms of a spacetime variational problem, with the microscopic properties of the system encoded in a membrane tension \cite{NahumEtAl2017,JonayHuseNahum2018,Mezei2018}.

The importance of such a description extends beyond the calculation of an entropy. Entanglement connects the characterization of quantum many-body states, the recovery of quantum information, and the emergence of spacetime in holography. The Ryu--Takayanagi and Hubeny--Rangamani--Takayanagi prescriptions express boundary entanglement in terms of bulk extremal surfaces \cite{RyuTakayanagi2006,HubenyRangamaniTakayanagi2007}, while the relation between entanglement and geometric connectivity \cite{VanRaamsdonk2010}, together with quantum-error-correcting interpretations of bulk locality and entanglement-wedge reconstruction \cite{AlmheiriDongHarlow2015,DongHarlowWall2016}, makes the geometric content of boundary information a central issue. Within this broader program, a concrete question is which parts of the bulk geometry are retained by a given effective description, and whether that information is sufficient to determine other observables that are not entanglement measures.

An important question raised by the membrane description concerns the amount of microscopic and geometric information retained by the effective tension. Although the membrane is formulated entirely in boundary spacetime, its tension is determined by the geometry of the gravitational dual. The reduction from a bulk extremal surface to a local membrane functional necessarily eliminates some of the original geometric data. It is therefore natural to ask whether the remaining information has an independent physical interpretation, whether it can be reconstructed from observables other than entanglement entropy, and which properties of the black-hole geometry remain inaccessible within the leading effective description. These questions are closely related to the broader problem of bulk reconstruction in holography, but here they can be addressed explicitly within a controlled classical gravitational setting. In particular, one may ask whether the geometric data encoded in the membrane tension also determine the response of an external probe that is governed by a different bulk variational principle.

The holographic membrane theory emerged from the study of entanglement growth after quenches. Hartman and Maldacena related the linear growth of entropy to extremal surfaces probing black-hole interiors \cite{HartmanMaldacena2013}. Liu and Suh identified universal regimes of holographic thermalization and developed the entanglement-tsunami picture \cite{LiuSuh2014Tsunami,LiuSuh2014Growth}. The subsequent analysis of large-region entanglement dynamics \cite{Mezei2017}, together with the minimal-cut and coarse-grained descriptions of chaotic many-body systems \cite{NahumEtAl2017,JonayHuseNahum2018}, led to an explicit holographic membrane formulation \cite{Mezei2018}. In this formulation, the bulk radial coordinate is eliminated in the leading large-region, late-time scaling limit, leaving a membrane in boundary spacetime with an orientation-dependent tension $\cE(u)$. Extensions to more general quench protocols, coupling to hydrodynamics, and higher-derivative corrections demonstrate the usefulness of this effective description beyond its simplest realization \cite{MezeiVirrueta2019}. Its application to reflected entropy, renormalization-group flows, and information velocity has further clarified both its scope and the special features of two-dimensional theories \cite{JiangMezeiVirrueta2025}.

A particularly useful feature of the trailing-string construction is that the velocity of the boundary quark controls the radial position of the induced worldsheet horizon. For a regular monotone black-brane geometry, this horizon approaches the spacetime horizon as the quark velocity tends to zero and moves toward the asymptotic boundary as the velocity approaches the speed of light. Consequently, varying the quark velocity provides a continuous way of probing the radial dependence of the bulk geometry through a boundary response function. The small-velocity expansion resolves the local structure near the black-hole horizon, whereas the complete finite-velocity force contains information accumulated over the exterior radial region. This observation suggests treating the drag force not only as a measure of momentum loss, but also as an inverse problem: to what extent can the geometry responsible for the force be reconstructed from its velocity dependence? The answer is not obvious, since the force is sensitive to only particular combinations of the bulk metric and possible string-frame couplings.

A different and extensively studied probe of strongly coupled dynamics is the drag force on a heavy external quark. In holography, stationary momentum loss is described by a trailing string whose endpoint moves through the thermal state at a prescribed velocity \cite{HerzogEtAl2006,Gubser2006}. This construction has been used to study heavy-probe energy loss in conformal plasmas and in nonconformal models motivated by QCD \cite{GursoyEtAl2009}. Fluctuations of the same string determine stochastic forces and momentum broadening \cite{GiecoldIancuMueller2009,GiataganasSoltanpanahi2014}. For review of holographic approach to QCD see \cite{Arefeva:2014QGP} and drag force relation to spatial string tension see\cite{Arefeva:2020QFT,Andreev:2017Drag, Sin:2006pv}. Unlike the entanglement membrane, the trailing string describes a continuously driven probe, and its mean force is fixed by regularity at a velocity-dependent worldsheet horizon. Nevertheless, the relation between the leading nonlinear drag correction and the butterfly velocity found in \cite{Ageev2021} suggests that stationary momentum loss can retain information about the geometric structures governing information propagation.

$\,$

In this work, we establish an exact correspondence between the finite-velocity drag force of a heavy quark and the geometric membrane tension governing holographic entanglement dynamics. We consider homogeneous and isotropic black-brane backgrounds with $p$ boundary spatial dimensions and a Nambu--Goto string probe with constant effective coupling. The stationary drag force can be written as $F(v)=-\sigma vH(v^2)$, where $v$ is the quark velocity, $\sigma$ is the linear drag coefficient, and $H(0)=1$. On the entanglement side, the membrane theory is characterized by a tension $\cE(u)$, which depends on the local membrane slope $u$. We show that these two functions are related exactly, without expanding in the quark velocity. In particular, for each velocity $v$, there is a corresponding point $u$ on the geometric membrane curve, determined parametrically by
\begin{equation}
v^2
=
u^2-\frac{u\cE(u)}{\cE'(u)},
\qquad
H(v^2)
=
\left[
\frac{\cE(u)\cE'(u)}{u}
\right]^{1/p}.
\label{eq:introexacttransform}
\end{equation}
Here the prime denotes differentiation with respect to the membrane slope $u$. The correspondence follows from the observation that the trailing-string horizon and the radial extremization underlying the membrane tension probe the same relation between the temporal and spatial warp factors of the bulk geometry. Importantly, the quark velocity $v$ and the membrane slope $u$ are distinct quantities. The limit $v\to0$ corresponds to $u\to v_B$, where $v_B$ is the butterfly velocity, while nonzero quark velocities probe the geometric continuation of the membrane beyond this point. The resulting transformation is invertible on a regular convex branch and provides a direct reconstruction of the exterior geometric membrane from the complete normalized drag curve.

The correspondence reveals a remarkable unity between two seemingly different aspects of strongly coupled dynamics. The stationary drag force describes the dissipative response of a heavy external probe and is determined holographically by the momentum flux along a trailing string, whereas the entanglement membrane governs the coarse-grained spreading of quantum entanglement and emerges from the large-scale reduction of bulk extremal surfaces. Despite their distinct physical origins and variational principles, both constructions are controlled by the same underlying  geometry. This shared geometric structure allows the membrane tension and the finite-velocity drag force to be related through an exact and invertible transformation, providing complementary descriptions of the same bulk information. The resulting correspondence establishes a direct bridge between quantum information dynamics and dissipative transport, and opens a systematic route to reconstructing holographic geometry from observables associated with different sectors of the boundary theory.

The small-velocity expansion resolves the local content of this relation. Writing
\begin{equation}
F(v)=-\sigma\left[v+c_1v^3+c_2v^5+c_3v^7+\cdots\right],
\label{eq:introdragseries}
\end{equation}
we find
\begin{equation}
\begin{aligned}
c_1&=\frac{1}{pv_B^2},\\
c_2&=\frac{p+1}{2p^2v_B^4}
-\frac{1}{pv_B^5\cE''(v_B)}.
\end{aligned}
\label{eq:introdragcoefficients}
\end{equation}
The first identity recovers the drag--butterfly relation of \cite{Ageev2021}. The second identifies the coefficient of $v^5$ as the first correction sensitive to the curvature of the membrane tension at its butterfly point. More generally, we derive an all-order inversion formula in which, for $n\geq2$, $c_n$ first introduces the $n$th derivative of the membrane tension. The nonlinear response thus supplies a hierarchy of local geometric information beyond $v_B$. The same structure also distinguishes the physical interior branch of the membrane from its exterior continuation: nonzero-velocity drag probes the latter, and its reconstruction requires the unmodified geometric tension rather than a linear extension adopted for entropy minimization.

An independent characterization of this horizon geometry is provided by scalar pole-skipping. At the frequencies $\omega_n=-in\kappa$, with $\kappa=2\pi T$, special complex momenta make the ingoing scalar horizon recursion degenerate \cite{BlakeDavisonVegh2020}. We combine the algebraic metric reconstruction from these degeneracies \cite{LuRanWu2026} with the membrane--drag transformation to express nonlinear momentum loss directly in terms of scalar spectral data. For a minimally coupled massless scalar, let $\mu_{n,a}=k_{n,a}^2$, and define $\epsilon=\mu_{1,1}$, $S_2=\mu_{2,1}+\mu_{2,2}$, and $P_2=\mu_{2,1}\mu_{2,2}$. With the area-radius horizon normalized to $r_h=1$, the first two levels give
\begin{equation}
\begin{aligned}
c_1&=-\frac{\epsilon}{p\kappa^2},\\
c_2&=\frac{
4p\epsilon S_2-pP_2-8(p-1)\epsilon^2
}{16p^2\kappa^4}.
\end{aligned}
\label{eq:intropoleskippingdrag}
\end{equation}
The tower through level $n$ determines the drag expansion through order $v^{2n+1}$ and the corresponding membrane endpoint derivatives. This establishes an explicit relation between degeneracies of a linear scalar fluctuation equation and the nonlinear response of a heavy probe. The first scalar skipped momentum also coincides with the momentum scale of the chaos-related energy-density pole-skipping point, although the two degeneracies occur at different frequencies and should not be identified as the same mode \cite{BlakeDavisonGrozdanovLiu2018}.

The nonlinear drag coefficient hierarchy also admits an interesting
application to the propagation and recovery of quantum information.
In particular, one may consider a strip-shaped information-recovery
protocol, in which the information velocity $v_I(f)$ characterizes
the asymptotic growth of the spatial region required to recover
initially localized quantum information
\cite{CouchEtAl2020,EcclesEtAl2021,JiangMezeiVirrueta2025}.
Here $f$ denotes the fraction of the final thermal entropy density
already present as initial volume-law entanglement. Within the
membrane description, this fraction enters through the
initial-state contribution to the variational problem, while
the membrane tension itself is determined by the final thermal
geometry. The resulting information velocity interpolates between
$v_I(0)=v_E$ and $v_I(1)=v_B$. This provides a natural setting
in which to investigate the information-theoretic consequences
of the geometric data encoded in the stationary drag force.
We first consider $f$ formally as a parameter entering the membrane
variational problem, allowing us to establish its relation to the
nonlinear drag coefficients within a fixed thermal geometry.
We then discuss possible physical interpretations and realizations
of this parameter in the heavy-quark setup, particularly through
drag-induced energy deposition and the associated entanglement deficit.

\begin{equation}
v_I(f)=v_B
-\frac{pv_B^5}{2}
\left[\frac{p+1}{2}c_1^2-c_2\right](1-f)
+O\!\left((1-f)^2\right).
\label{eq:introinformationdrag}
\end{equation}
Thus the quintic drag coefficient fixes the leading departure of the information-recovery velocity from the butterfly velocity. Higher drag coefficients determine successive corrections, provided the interior and exterior membrane branches are smooth at their common horizon endpoint.

Conversely, the full information-velocity function supplies a reconstruction of the physical interior branch without analytically continuing exterior drag data. For a smooth strictly convex strip membrane associated with a fixed final thermal geometry, define $u_f=v_I(f)-(1-f)v_I'(f)$, where the prime now denotes differentiation with respect to $f$. We find
\begin{equation}
\begin{aligned}
Y(f)&=\frac{f\left[v_I(f)-f(1-f)v_I'(f)\right]}{u_f},\\
a(f)&=-\frac{u_f(1-f)v_I(f)}{f},
\qquad 0<f<1.
\end{aligned}
\label{eq:introinteriorreconstruction}
\end{equation}
These expressions determine $a(Y)$ from the horizon to the entanglement bottleneck. Physical drag and information recovery therefore reconstruct complementary real intervals of the same reduced geometry. The local relation \eqref{eq:introinformationdrag} describes their matching at the horizon; it does not imply that a finite set of exterior drag coefficients fixes the entire interior recovery curve.

The remaining results clarify the constraints and limitations of this reconstruction. In Einstein gravity with matter satisfying the null energy condition, a nonnegative causal-load function organizes bounds on the membrane, information velocities, and normalized drag. In particular, $H(x)\geq(1-x)^{-2/(p+1)}$ at fixed horizon normalization, with AdS--Schwarzschild as the reference geometry saturating the bound. A radially varying string coupling introduces independent probe data and modifies the geometric force relations. Moreover, the membrane and mean drag alone leave the radial proper-distance element undetermined. We show that this missing function is fixed by one absolute momentum-broadening curve: the relation $T_{\mathrm{ws}}(v)=\kappa_\perp(v)/[2\sigma H(v^2)]$ determines the worldsheet temperature, from which the full exterior metric follows for a common-frame probe, up to an overall normalization. The distinction between reduced geometry, radial propagation, and probe coupling is essential to identifying precisely what each boundary observable reconstructs.

The paper is organized as follows. In \cref{sec:setup,sec:exacttransform}, we derive the two variational problems and their exact tangent transformation. Section~\ref{sec:coefficients} develops the nonlinear coefficient hierarchy, while \cref{sec:NEC,sec:dressing} discuss null-energy constraints and probe-frame dependence. The exact examples in \cref{sec:examples} illustrate the Schwarzschild and charged geometries, including the nonanalytic extremal limit. Section~\ref{sec:poleskipping} relates the scalar pole-skipping tower to drag and membrane data. Information recovery and interior reconstruction are developed in \cref{sec:informationvelocity}, and \cref{sec:stochasticreconstruction} completes the exterior metric using stochastic fluctuations. Further implications and special cases are discussed in \cref{sec:further}, before the conclusions in \cref{sec:conclusions}.

\section{Reminder on drag force and entanglement membrane}
\label{sec:setup}

In general we consider a homogeneous and isotropic thermal state of a holographic field theory in \(p+1\) spacetime dimensions, whose gravitational dual is described by a \((p+2)\)-dimensional black-brane geometry. At this stage, we do not assume any particular bulk gravitational theory. As we will see, several of the relations derived below are purely geometric and do not rely on the specific form of the gravitational field equations.

To keep the discussion independent of any particular choice of radial coordinate, we write the metric\footnote{In general we assume Einstein frame here but in what follows we also consider string frame.} in the general form
\begin{equation}
\dd s_E^2
=
-D(r)\,\dd t^2
+\frac{\dd r^2}{K(r)}
+C(r)\,\delta_{ij}\dd x^i\dd x^j,
\qquad
i=1,\ldots,p ,
\label{eq:generalmetric}
\end{equation}
where the horizon is located at \(r=r_h\), and the radial coordinate increases toward the asymptotic boundary.

Two combinations of the metric functions will play a central role throughout our analysis. The first is
\begin{equation}
a(r)
=
\frac{D(r)}{C(r)} ,
\label{eq:adef}
\end{equation}
which characterizes the local causal structure through the relative redshift between the temporal and spatial directions. For a regular outer horizon, \(a(r_h)=0\), while the boundary coordinates are normalized such that \(a(r)\to 1\) asymptotically.

The second is the dimensionless transverse-area variable
\begin{equation}
Y(r)
=
\left(\frac{C(r)}{C_h}\right)^p,
\qquad
C_h=C(r_h),
\label{eq:Ydef}
\end{equation}
which measures the squared transverse area density relative to its horizon value. By construction, \(Y(r_h)=1\), and for the geometries considered below it increases monotonically toward the boundary. Together, \(a(r)\) and \(Y(r)\) provide a convenient characterization of the bulk geometry that does not depend on the choice of radial coordinate. We refer to the radial-coordinate-independent relation \(a=a(Y)\) as the causal-area relation, since it connects the local causal structure of the geometry to its transverse-area density.

We further introduce the nonnegative function \(\mathcal N(x)\), which we call the causal load. It characterizes the departure of the reduced geometry from AdS–Schwarzschild and encodes the contributions of matter satisfying the null energy condition.

$\,$

We now turn to two seemingly distinct observables: the drag force experienced by a heavy quark moving through the thermal medium and the coarse-grained entanglement dynamics of a large boundary region. In the holographic description, the former is determined by a classical trailing string whose endpoint moves at a fixed velocity, while the latter is governed by the scaling limit of the Hubeny--Rangamani--Takayanagi (HRT) surface \cite{RyuTakayanagi2006,HubenyRangamaniTakayanagi2007,Mezei2018}.

Although these observables arise from different variational principles, they share a common geometric structure. In both cases, the radial coordinate can be eliminated algebraically, leaving a description entirely in terms of the same reduced geometric data. For the trailing string, this reduction determines the conserved momentum flux through a velocity-dependent worldsheet horizon. For the entanglement membrane, it determines the effective tension associated with a given local spacetime orientation.

As we will show, these two constructions are connected through the causal-area relation \(a(Y)\) introduced above. The resulting correspondence does not identify the drag force with entanglement entropy; rather, it establishes that their respective geometric descriptions are related by an exact transformation, with both probing the same underlying bulk geometry in complementary ways.

\subsection{Trailing string and exact momentum flux}
\label{subsec:trailing}

We begin with the holographic description of a heavy quark moving at constant velocity through the thermal medium. In the bulk, the quark is represented by the endpoint of an open string extending from the asymptotic boundary toward the black-brane horizon. Working in Einstein-frame variables, we consider the string action
\begin{equation}
S_{\mathrm{str}}
=
-T_f
\int \dd^2\sigma\,
Z(r)\sqrt{-\det h^{(E)}_{ab}},
\qquad
T_f=\frac{1}{2\pi\alpha'} ,
\label{eq:stringaction}
\end{equation}
where \(h^{(E)}_{ab}\) is the induced Einstein-frame worldsheet metric and \(Z(r)>0\) accounts for the radial dependence of the effective string tension, including possible scalar-dependent transformations between the Einstein and string frames. The case of constant \(Z\) will be of particular interest, as it allows the normalized drag force to be determined entirely by the same geometric data that govern the entanglement membrane. For the moment, however, we keep \(Z(r)\) arbitrary to distinguish the universal geometric contribution from possible probe-dependent effects.

Choosing the static gauge with worldsheet coordinates \((t,r)\), we describe a quark moving along the \(x^1\) direction with constant velocity \(v\) through the stationary embedding
\begin{equation}
x^1(t,r)
=
vt+\xi(r),
\qquad
x^{i>1}=0 .
\label{eq:trailingansatz}
\end{equation}
The function \(\xi(r)\) describes the trailing profile of the string induced by its interaction with the thermal medium. Substituting this ansatz into the background metric \eqref{eq:generalmetric}, the nonvanishing components of the induced worldsheet metric are
\begin{equation}
h_{tt}^{(E)}
=
-D+Cv^2,
\qquad
h_{tr}^{(E)}
=
Cv\xi',
\qquad
h_{rr}^{(E)}
=
\frac{1}{K}+C\xi'^2 ,
\label{eq:inducedmetric}
\end{equation}
where a prime denotes differentiation with respect to \(r\). The corresponding determinant takes the form
\begin{equation}
-\det h^{(E)}
=
\frac{D-Cv^2}{K}
+DC\,\xi'^2 .
\label{eq:detinduced}
\end{equation}

Since the string action depends on \(\xi\) only through its radial derivative, the associated conjugate momentum is conserved along the string. Explicitly,
\begin{equation}
\Pi_\xi
=
\frac{ZDC\,\xi'}
{\sqrt{(D-Cv^2)/K+DC\,\xi'^2}} .
\label{eq:pixi}
\end{equation}
This conserved quantity determines the radial transport of spatial momentum and, consequently, the drag force acting on the boundary quark. Solving for the string profile in terms of \(\Pi_\xi\), we obtain
\begin{equation}
\xi'^2
=
\frac{
\Pi_\xi^2(D-Cv^2)
}{
KDC\left(Z^2DC-\Pi_\xi^2\right)
}.
\label{eq:xiprime}
\end{equation}

The physical solution is fixed by requiring the string profile to remain real and regular throughout the exterior geometry. In particular, the numerator vanishes at a radial position \(r=r_s(v)\) satisfying \(D(r_s)=v^2C(r_s)\). This is precisely the location where the induced worldsheet metric develops a horizon. Using the definition \eqref{eq:adef}, the worldsheet-horizon condition becomes
\begin{equation}
a(r_s)
=
v^2 .
\label{eq:worldsheethorizon}
\end{equation}
Regularity requires the denominator of \eqref{eq:xiprime} to vanish at the same radial position, thereby fixing the conserved momentum flux to be
\begin{equation}
\Pi_\xi
=
v\,Z(r_s)C(r_s) .
\label{eq:fluxregularity}
\end{equation}
The momentum transferred from the moving quark to the thermal medium is carried down the string toward the black-brane horizon. The resulting drag force is therefore
\begin{equation}
F(v)
=
-T_f\,v\,Z(r_s)C(r_s),
\qquad
a(r_s)=v^2 .
\label{eq:exactdrag}
\end{equation}
This is the standard trailing-string result \cite{HerzogEtAl2006,Gubser2006}, expressed without fixing the radial coordinate or specifying the matter content supporting the background geometry.

An important feature of \eqref{eq:exactdrag} is that the drag force depends only on the local values of the transverse warp factor \(C(r)\) and the effective string coupling \(Z(r)\) at the worldsheet horizon, whose position is determined by \(a(r_s)=v^2\). In particular, the radial metric function \(K(r)\) drops out of the final expression. Although \(K(r)\) influences the detailed shape of the trailing string, it does not enter the conserved momentum flux. This distinction will be central to the connection with the entanglement membrane developed below.

To isolate the nontrivial velocity dependence, we introduce the horizon normalization
\begin{equation}
\sigma
=
T_f Z_h C_h,
\qquad
Z_h=Z(r_h),
\label{eq:sigmadef}
\end{equation}
and express the drag force as
\begin{equation}
F(v)
=
-\sigma v\,H_Z(v^2),
\qquad
H_Z(0)=1 .
\label{eq:Hdef}
\end{equation}
The dimensionless function \(H_Z\) then takes the exact form
\begin{equation}
H_Z(v^2)
=
\frac{Z(r_s)}{Z_h}
\frac{C(r_s)}{C_h}.
\label{eq:HZradial}
\end{equation}
The normalization follows from the fact that the worldsheet horizon approaches the spacetime horizon as \(v\to0\), so that \(r_s(0)=r_h\).

The overall force scale can also be related directly to the thermodynamic properties of the black brane. In two-derivative Einstein gravity, the thermal entropy density is given by
\(s_{\mathrm{th}}=C_h^{p/2}/(4G_N)\), which implies
\begin{equation}
\sigma
=
T_f Z_h
\left(4G_Ns_{\mathrm{th}}\right)^{2/p}.
\label{eq:sigmaentropy}
\end{equation}
Thus, the normalization of the drag force is controlled jointly by the horizon area density and the coupling of the string probe. Its remaining velocity dependence is encoded in \(H_Z(v^2)\), which measures how the effective string tension and transverse geometry vary as the worldsheet horizon moves away from the spacetime horizon. In the constant-\(Z\) case, this dependence is purely geometric and is determined by the relation between the local causal structure \(a(r)\) and the transverse-area variable \(Y(r)\). As we will see, this is precisely the geometric information captured by the entanglement membrane.

\subsection{The HRT scaling limit and the geometric membrane}
\label{subsec:hrtmembrane}

We now turn to the holographic entanglement membrane and derive its geometric tension using the same radial variables introduced above. The starting point is the HRT prescription, in which the entanglement entropy of a boundary region is determined by the area of an extremal codimension-two surface in the bulk. To describe surfaces that extend across the black-brane horizon, it is convenient to introduce the ingoing Eddington--Finkelstein coordinate
\begin{equation}
\tau
=
t+\int^r\frac{\dd \bar r}{\sqrt{D(\bar r)K(\bar r)}} ,
\label{eq:ingoingtime}
\end{equation}
in terms of which the Einstein-frame metric becomes
\begin{equation}
\dd s_E^2
=
-D\,\dd\tau^2
+2\sqrt{\frac{D}{K}}\,\dd\tau\,\dd r
+C\,\dd\bm{x}_p^2 .
\label{eq:ingoingmetric}
\end{equation}
This coordinate system is regular at the future horizon and provides a natural setting for analyzing the large-scale dynamics of extremal surfaces.

Locally, we parametrize the HRT surface by
\begin{equation}
x^1=X(\tau,\bm y),
\qquad
r=r(\tau,\bm y),
\label{eq:HRTembedding}
\end{equation}
where \(\bm y\) denotes the remaining \(p-1\) boundary spatial coordinates. The membrane description emerges in the scaling limit appropriate to large entangling regions, in which the boundary spatial coordinates and time are rescaled by a common large parameter, while the bulk radial coordinate remains unscaled. In this limit, derivatives of the radial embedding are parametrically suppressed, and the leading contribution to the induced area density simplifies to
\begin{equation}
\sqrt{\det h}
=
C^{p/2}
\sqrt{1+\lvert\nabla X\rvert^2}
\sqrt{u^2-a(r)} ,
\label{eq:HRTleading}
\end{equation}
where
\begin{equation}
u
=
\frac{\partial_\tau X}
{\sqrt{1+\lvert\nabla X\rvert^2}}
\label{eq:uslope}
\end{equation}
defines the local membrane slope. The dependence on the bulk geometry is therefore reduced to the transverse warp factor \(C(r)\) and the local causal function \(a(r)\).

Normalizing the area density by the thermal entropy density and using the transverse-area variable \(Y(r)\) defined in \eqref{eq:Ydef}, we obtain the squared local off-shell membrane tension,
\begin{equation}
\mathscr{L}(r,u)
=
Y(r)\left[u^2-a(r)\right].
\label{eq:offshellmembrane}
\end{equation}
A key feature of the scaling limit is that the radial coordinate no longer appears through its derivatives. Instead, it becomes an auxiliary variable whose value is determined locally by extremizing the area density at fixed membrane slope. The corresponding stationarity condition is
\begin{equation}
Y'(r)\left[u^2-a(r)\right]
-
Y(r)a'(r)
=
0 ,
\label{eq:radialsaddle}
\end{equation}
which can be rearranged into
\begin{equation}
u^2
=
a(r)
+
\frac{Y(r)a'(r)}{Y'(r)}
=
a(r)
+
\frac{C(r)a'(r)}{pC'(r)} .
\label{eq:uofr}
\end{equation}
This relation determines the radial position associated with a given membrane slope. Substituting the resulting solution \(r=r(u)\) back into the area density yields the geometric membrane tension,
\begin{equation}
\cE(u)^2
=
Y(r(u))
\left[u^2-a(r(u))\right].
\label{eq:Eonshell}
\end{equation}

An important consequence of this reduction is that the radial metric function \(K(r)\) drops out of both the extremization condition and the resulting membrane tension. As in the trailing-string calculation, the relevant geometric information is therefore encoded entirely in the relation between \(a(r)\) and \(Y(r)\). This common dependence provides the basis for the exact correspondence between the membrane tension and the stationary drag force developed in the following section.

The connection becomes particularly transparent at the black-brane horizon. Evaluating \eqref{eq:uofr} at \(r=r_h\), where \(a(r_h)=0\), identifies the corresponding membrane slope with the butterfly velocity,
\begin{equation}
v_B^2
=
\frac{C_h a'_h}{pC'_h}
=
\frac{D'_h}{pC'_h},
\label{eq:vB}
\end{equation}
in agreement with the standard result obtained from the near-horizon shockwave analysis. The horizon thus defines a distinguished point of the membrane tension, at which
\begin{equation}
\cE(v_B)=v_B,
\qquad
\cE'(v_B)=1 .
\label{eq:endpointconditions}
\end{equation}
The first identity follows immediately from \eqref{eq:Eonshell}, using \(Y(r_h)=1\) and \(a(r_h)=0\). The second follows by differentiating the on-shell tension with respect to \(u\): because the radial embedding satisfies the stationarity condition \eqref{eq:radialsaddle}, its implicit variation does not contribute.

These endpoint relations will play a central role in the analysis of the velocity expansion of the drag force. In particular, they imply that the leading nonlinear correction, proportional to \(v^3\), is determined entirely by the butterfly velocity, while successive higher-order terms probe the curvature and higher derivatives of the membrane tension at \(u=v_B\).

\section{Exact membrane representation of the drag force}
\label{sec:exacttransform}

The preceding analysis shows that both the trailing-string dynamics and the entanglement membrane are governed by the same reduced geometric data, encoded in the functions \(a(r)\) and \(Y(r)\). We now make this connection explicit by showing that the radial extremization underlying the membrane construction admits a natural Legendre-transform structure, which in turn provides an exact representation of the drag force.

To expose this structure, it is convenient to introduce the squared membrane slope
\begin{equation}
q=u^2
\label{eq:qdef}
\end{equation}
and regard the squared membrane tension as a function of \(q\),
\begin{equation}
M(q)
=
\cE(\sqrt q)^2 .
\label{eq:Mdef}
\end{equation}
Since the transverse-area variable \(Y\) provides a radial parametrization of the geometry, we may also express the local causal function as \(a=a(Y)\) and define
\begin{equation}
G(Y)
=
Y\,a(Y).
\label{eq:Gdef}
\end{equation}
In terms of these variables, the radial stationarity condition \eqref{eq:radialsaddle} takes the simple form
\begin{equation}
q
=
\frac{\dd G}{\dd Y},
\label{eq:qGprime}
\end{equation}
while the on-shell membrane tension becomes
\begin{equation}
M(q)
=
qY-G(Y).
\label{eq:Legendre}
\end{equation}
These two relations reveal that \(M(q)\) and \(G(Y)\) are related by a Legendre transformation, understood locally on each smooth branch of the radial saddle. The membrane tension therefore provides an alternative description of the same reduced bulk geometry, with the squared slope \(q\) playing the role of the variable conjugate to the transverse-area density \(Y\).

This correspondence can be made more explicit by differentiating \eqref{eq:Legendre} with respect to \(q\). Using the stationarity condition \eqref{eq:qGprime}, the implicit dependence of \(Y\) cancels, leaving the envelope relation
\begin{equation}
M'(q)
=
Y.
\label{eq:MprimeY}
\end{equation}
Combining this identity with \eqref{eq:Legendre}, we can reconstruct the local causal function directly from the membrane tension,
\begin{equation}
a(q)
=
q-\frac{M(q)}{M'(q)} .
\label{eq:afromM}
\end{equation}
Equations \eqref{eq:MprimeY} and \eqref{eq:afromM} thus determine both \(Y\) and \(a\) parametrically in terms of \(q\). In particular, the unmodified geometric membrane tension contains sufficient information to reconstruct the causal-area relation \(a(Y)\) on the corresponding branch. This reconstruction does not, however, determine the radial lapse, which drops out of the membrane construction. As a result, the membrane captures the geometric information relevant to the stationary drag force, but not necessarily that required by observables involving radial propagation or integration through the bulk.

We can now establish the connection with the trailing-string result. Introducing the squared quark velocity
\begin{equation}
x=v^2
\label{eq:xdef}
\end{equation}
allows us to distinguish the physical velocity \(v\) from the auxiliary membrane slope \(u\). Recall that the worldsheet horizon is determined by the condition \(a(r_s)=v^2\). Using \eqref{eq:afromM}, this condition selects a point \(q=q(x)\) on the exterior branch of the geometric membrane through the implicit relation
\begin{equation}
x
=
q-\frac{M(q)}{M'(q)} .
\label{eq:xqrelation}
\end{equation}
The quark velocity therefore determines which point of the membrane curve is probed by the worldsheet horizon.

For a constant effective string coupling \(Z\), the normalized drag force depends only on the transverse warp factor evaluated at this horizon,
\(H(x)=C(r_s)/C_h\). Recalling that \(Y=(C/C_h)^p\), and using the Legendre relation \(M'(q)=Y\), we arrive at
\begin{equation}
H(x)
=
M'(q(x))^{1/p}.
\label{eq:Hexactparametric}
\end{equation}
Together, \eqref{eq:xqrelation} and \eqref{eq:Hexactparametric} provide an exact parametric representation of the normalized drag force in terms of the geometric membrane tension.

Importantly, this relation has been obtained without expanding in the quark velocity. The finite-velocity drag force is determined by the derivative of the squared membrane tension, evaluated at the point where the reconstructed causal function equals \(v^2\). The membrane tension and the drag force are therefore not independent geometric quantities: they encode the same causal-area relation \(a(Y)\), expressed through two different prescriptions for eliminating the bulk radial coordinate.

\subsection{Tangent transform and its inverse}
\label{subsec:tangent}

The exact parametric relation derived above admits a particularly simple geometric interpretation. Rearranging \eqref{eq:xqrelation}, we obtain
\begin{equation}
M(q)
=
M'(q)(q-x).
\label{eq:tangencycondition}
\end{equation}
This equation has a direct interpretation in the \((q,M)\)-plane. At the point selected by the quark velocity, the tangent line to the membrane curve has slope \(M'(q)\) and intersects the horizontal axis at \(q=x\). Thus, the squared quark velocity determines the horizontal intercept of the tangent, while its slope is precisely the \(p\)-th power of the normalized drag force, \(M'(q)=H(x)^p\).

This geometric construction can be expressed as a variational transformation. When \(M(q)\) is convex on the relevant exterior branch, the tangent condition selects the minimum of the ratio \(M(q)/(q-x)\), giving
\begin{equation}
H(x)^p
=
\min_{q>x}
\frac{M(q)}{q-x}.
\label{eq:forwardtransform}
\end{equation}
Indeed, extremizing the right-hand side with respect to \(q\) reproduces \eqref{eq:tangencycondition}, while the convexity condition \(M''(q)>0\) ensures that the stationary point is a minimum. The normalized drag force can therefore be obtained directly from the membrane tension through a geometric optimization, without explicitly reconstructing the bulk radial coordinate.

Remarkably, this transformation is invertible. To make the inverse construction transparent, we introduce
\begin{equation}
\mathcal{Y}(x)
=
H(x)^p .
\label{eq:calYdef}
\end{equation}
Each value of \(x\) determines a tangent line with slope \(\mathcal{Y}(x)\) and horizontal intercept \(x\). At its point of contact with the membrane curve, \(q=q(x)\), the tangent relation reads
\begin{equation}
M(q)
=
\mathcal{Y}(x)(q-x).
\label{eq:tangentfamily}
\end{equation}
The membrane curve can then be reconstructed as the envelope of this one-parameter family of tangent lines. Imposing the envelope condition by differentiating with respect to \(x\), at fixed \(q\), gives
\begin{equation}
q(x)
=
x+\frac{\mathcal{Y}(x)}{\mathcal{Y}'(x)}
=
x+\frac{H(x)}{pH'(x)}.
\label{eq:qfromH}
\end{equation}
Substituting this result into the tangent relation yields
\begin{equation}
M(q(x))
=
\frac{\mathcal{Y}(x)^2}{\mathcal{Y}'(x)}
=
\frac{H(x)^{p+1}}{pH'(x)} .
\label{eq:MfromH}
\end{equation}
These expressions provide an explicit parametric reconstruction of the exterior membrane tension from the complete normalized drag curve. Conversely, the envelope construction can be written in the variational form
\begin{equation}
M(q)
=
\sup_{x<q}
\mathcal{Y}(x)(q-x).
\label{eq:inversetransform}
\end{equation}
Together, \eqref{eq:forwardtransform} and \eqref{eq:inversetransform} establish a pair of mutually inverse transformations between the geometric membrane tension and the normalized drag force, on the corresponding regular convex branch. Their structure is closely related to Legendre--Fenchel duality, with a distinctive geometric feature: the physical variable \(x\) labels the horizontal intercept of a tangent line, rather than its slope. In this sense, the correspondence can be viewed as a projective form of Legendre duality.

Beyond relating two observables, this correspondence also provides a direct reconstruction of the reduced bulk geometry. Indeed, the drag force determines the transverse-area variable evaluated at the worldsheet horizon,
\begin{equation}
Y(x)
=
H(x)^p,
\qquad
a=x ,
\label{eq:reducedtomography}
\end{equation}
where the second relation follows from the worldsheet-horizon condition. Consequently, knowledge of the normalized drag force over the full velocity range determines the causal-area relation \(a(Y)\) throughout the corresponding exterior region. Using \(Y\) itself as the radial coordinate, the Einstein-frame metric can then be expressed as
\begin{equation}
\dd s_E^2
=
C_hY^{1/p}
\left[-a(Y)\,\dd t^2+\dd\bm{x}_p^2\right]
+\mathcal{B}_Y(Y)\,\dd Y^2 .
\label{eq:reducedmetric}
\end{equation}
The function \(\mathcal{B}_Y(Y)\), which controls the radial proper-distance element, remains undetermined by this construction. The reconstruction therefore fixes the temporal and spatial warp factors relative to the horizon scale, but not the complete bulk metric.

This limitation also clarifies the physical scope of the correspondence. Both the stationary drag force and the geometric membrane tension depend only on the causal-area relation \(a(Y)\), whereas observables involving radial propagation or integration generally retain sensitivity to the radial lapse. Consequently, the exact membrane--drag correspondence does not, by itself, determine quantities such as the generic quasinormal-mode spectrum, absolute Langevin coefficients, screening lengths, or radial Wilson-loop integrals. Rather, it identifies the precise sector of bulk geometric information that is shared by the two variational problems.

\subsection{The butterfly seam and the two membrane branches}
\label{subsec:seam}

The correspondence between the membrane tension and the drag force acquires a particularly interesting structure at the black-brane horizon. In the zero-velocity limit, \(x=v^2\to0\), the worldsheet horizon approaches the spacetime horizon, and the associated membrane slope approaches the butterfly velocity. The corresponding point on the squared membrane-tension curve is therefore
\begin{equation}
q(0)
=
v_B^2
\equiv B,
\label{eq:Bdef}
\end{equation}
where the horizon endpoint conditions \eqref{eq:endpointconditions} imply
\begin{equation}
M(B)
=
B,
\qquad
M'(B)
=
1 .
\label{eq:Mendpoint}
\end{equation}
These relations have a simple geometric interpretation. The tangent line to \(M(q)\) at \(q=B\) has unit slope and passes through the origin, so that its equation is \(M=q\). The butterfly velocity thus identifies a distinguished point on the membrane curve: the point of tangency with the unit-slope line passing through the origin. In this way, the near-horizon butterfly velocity acquires a direct interpretation in the tangent construction underlying the membrane--drag correspondence.

An important aspect of this correspondence is that the two observables probe different branches of the same geometric membrane. The minimal membranes governing the large-scale growth of entanglement typically explore the range
\begin{equation}
0\leq q\leq v_B^2,
\label{eq:interiorbranch}
\end{equation}
which corresponds to the black-hole interior, extending from the horizon toward the entanglement bottleneck. By contrast, a quark moving with nonzero velocity probes the exterior branch,
\begin{equation}
v_B^2<q<1 ,
\label{eq:exteriorbranch}
\end{equation}
as its worldsheet horizon moves outward from the spacetime horizon toward the asymptotic boundary. The point \(q=v_B^2\) therefore marks the junction between the interior and exterior branches, providing the common geometric reference point for entanglement dynamics and stationary drag.

This distinction is important for the reconstruction developed above. While the interior branch is the one ordinarily relevant for the minimization of the membrane action, the exact finite-velocity drag force requires knowledge of the geometric membrane tension on the exterior branch. Consequently, one must retain the original tension obtained from the radial extremization, including its continuation beyond the butterfly point. An effective membrane tension modified by convexification or linear extension for the purposes of entropy minimization need not preserve the exterior information required to reconstruct the drag force.

The parametric correspondence also makes clear that the membrane slope \(u\) should not be identified with the physical quark velocity \(v\). Indeed, combining the Legendre relations with the worldsheet-horizon condition gives
\begin{equation}
q-x
=
\frac{M(q)}{M'(q)}
=
\frac{\cE(u)^2}{Y},
\label{eq:qxpositive}
\end{equation}
which is strictly positive along a regular branch with positive membrane tension. Since \(q=u^2\) and \(x=v^2\), the membrane slope associated with a given quark velocity satisfies \(u>v\). In particular, the membrane slope remains finite as the quark velocity vanishes, approaching \(u=v_B\) rather than zero.

Near this horizon endpoint, the relation between the two variables can be expanded systematically. Using the endpoint conditions \eqref{eq:Mendpoint}, one finds
\begin{equation}
u(v)
=
v_B
+
\frac{v^2}{v_B^2\cE''(v_B)}
+O(v^4).
\label{eq:uexpansion}
\end{equation}
This expansion shows that the leading departure of the membrane slope from the butterfly point is controlled by the local curvature of the membrane tension. It provides the link between the small-velocity expansion of the drag force and the successive derivatives of \(\cE(u)\) at \(u=v_B\), which will be developed in the following section.

\section{Velocity expansion and the membrane endpoint jet}
\label{sec:coefficients}

Although the exact membrane--drag transformation determines the force at arbitrary velocity, its small-velocity expansion provides a complementary perspective by resolving the local geometric structure of the membrane near the butterfly point. In particular, it establishes a systematic relation between the nonlinear corrections to the drag force and the successive derivatives of the membrane tension at \(u=v_B\).

To make this relation explicit, we expand the normalized drag function in powers of the squared quark velocity,
\begin{equation}
H(x)
=
1+\sum_{n=1}^{\infty}c_nx^n,
\qquad
x=v^2 ,
\label{eq:Hseries}
\end{equation}
so that the drag force takes the form
\begin{equation}
F(v)
=
-\sigma
\left(
v+c_1v^3+c_2v^5+c_3v^7+\cdots
\right).
\label{eq:Fseries}
\end{equation}
The relation between the leading nonlinear coefficient \(c_1\) and the butterfly velocity was previously identified in \cite{Ageev2021}. The exact transformation derived above places this result within a broader structure: the entire velocity expansion is organized into a triangular hierarchy, in which each successive coefficient probes an additional derivative of the geometric membrane tension at the butterfly point.

This hierarchy can be derived to arbitrary order using the Lagrange--B\"urmann inversion formula. To this end, we introduce the auxiliary function
\begin{equation}
N(q)
=
qM'(q)-M(q).
\label{eq:Ndef}
\end{equation}
The parametric relation \eqref{eq:xqrelation}, together with the envelope identity \(Y=M'(q)\), implies
\begin{equation}
N=xY .
\label{eq:Nimplicit}
\end{equation}
At the butterfly point \(q=B=v_B^2\), the endpoint conditions give \(N(B)=0\), making \(N\) a natural expansion variable near the horizon. Differentiating its definition, we find
\begin{equation}
\frac{\dd N}{\dd q}
=
qM''(q),
\qquad
\frac{\dd Y}{\dd N}
=
\frac{1}{q}.
\label{eq:Ndifferentials}
\end{equation}
The second identity is particularly useful because it expresses the variation of the transverse-area density directly in terms of the squared membrane slope.

Since \(N(x)\) is determined implicitly by \(N=xY(N)\), the Lagrange--B\"urmann formula gives, for an arbitrary differentiable function \(\Phi(N)\),
\begin{equation}
[x^n]\,\Phi(N(x))
=
\frac{1}{n}
[N^{n-1}]
\Phi'(N)Y(N)^n .
\label{eq:Lagrangeidentity}
\end{equation}
Here, \([x^n]\) denotes extraction of the coefficient of \(x^n\) in the corresponding series. Applying this identity to \(\Phi(N)=Y(N)^{1/p}\), and using \eqref{eq:Ndifferentials} to express derivatives with respect to \(N\) in terms of \(q\), we obtain the all-order result
\begin{equation}
c_n
=
\frac{1}{p\,n!}
\left.
\left(
\frac{1}{qM''(q)}
\frac{\dd}{\dd q}
\right)^{n-1}
\left[
\frac{M'(q)^{\,n-1+1/p}}{q}
\right]
\right|_{q=B},
\qquad
n\geq 1 .
\label{eq:allordercn}
\end{equation}
This formula expresses every coefficient in the nonlinear drag expansion in terms of the local derivatives of the squared membrane tension at the butterfly point.

The structure of \eqref{eq:allordercn} is especially revealing. For \(n\geq2\), the coefficient \(c_n\) involves derivatives of \(M(q)\) only up to order \(n\), or equivalently derivatives of \(\cE(u)\) up to the same order. Thus, increasing the order of the velocity expansion introduces precisely one new local geometric datum. This triangular dependence is a direct consequence of the envelope structure of the membrane transformation, which leads to the simplification \(\dd Y/\dd N=1/q\).

To display the first few coefficients explicitly, it is convenient to work with the causal-area function \(G(Y)=Ya(Y)\) introduced in \eqref{eq:Gdef}. At the horizon, \(Y=1\), and the endpoint conditions imply \(G(1)=0\) and \(G'(1)=B\). We define
\begin{equation}
\nu
=
\frac{1}{p},
\qquad
\Gamma_j
=
\frac{G^{(j)}(1)}{B},
\qquad
j\geq2 ,
\label{eq:Gammadef}
\end{equation}
where all derivatives are evaluated at the horizon. Expanding the relation \(x=G(Y)/Y\) around \(Y=1\) and inverting the resulting series gives
\begin{equation}
c_1
=
\frac{\nu}{B},
\label{eq:c1Gamma}
\end{equation}
\begin{equation}
c_2
=
\frac{\nu}{2B^2}
\left(
\nu+1-\Gamma_2
\right),
\label{eq:c2Gamma}
\end{equation}
\begin{equation}
c_3
=
\frac{\nu}{6B^3}
\left[
(\nu+1)(\nu+2)
-3(\nu+2)\Gamma_2
+3\Gamma_2^2
-\Gamma_3
\right],
\label{eq:c3Gamma}
\end{equation}
and
\begin{equation}
\begin{aligned}
c_4
=
\frac{\nu}{24B^4}
\big[
&(\nu+1)(\nu+2)(\nu+3)
-6(\nu+2)(\nu+3)\Gamma_2
\\
&+15(\nu+3)\Gamma_2^2
-15\Gamma_2^3
-4(\nu+3)\Gamma_3
\\
&+10\Gamma_2\Gamma_3
-\Gamma_4
\big].
\end{aligned}
\label{eq:c4Gamma}
\end{equation}
These expressions make the hierarchical structure explicit: \(c_1\) depends only on the first derivative of \(G\) at the horizon, while each subsequent coefficient introduces one additional derivative of the causal-area function.

The connection with the membrane tension follows directly from the Legendre relations. Since \(q=G'(Y)\) and \(Y=M'(q)\), the derivatives of the two functions are related by the standard inverse-function identities. In particular,
\begin{equation}
\Gamma_2
=
\frac{1}{B M''(B)},
\label{eq:Gamma2M}
\end{equation}
\begin{equation}
\Gamma_3
=
-\frac{M'''(B)}
{B M''(B)^3},
\label{eq:Gamma3M}
\end{equation}
and
\begin{equation}
\Gamma_4
=
\frac{
3M'''(B)^2-M''(B)M''''(B)
}{
B M''(B)^5
}.
\label{eq:Gamma4M}
\end{equation}
These relations allow the velocity coefficients to be expressed entirely in terms of derivatives of the membrane tension, rather than derivatives of the bulk metric functions.

To make this dependence more transparent, let \(b=v_B\) and introduce the shorthand \(e_j=\cE^{(j)}(b)\). Using \(M(q)=\cE(\sqrt q)^2\), together with the endpoint conditions \(\cE(b)=b\) and \(\cE'(b)=1\), we obtain
\begin{equation}
M''(B)
=
\frac{e_2}{2b},
\qquad
M'''(B)
=
\frac{e_3}{4b^2},
\label{eq:MderivativesEfirst}
\end{equation}
and
\begin{equation}
M''''(B)
=
\frac{e_4}{8b^3}
-\frac{e_3}{4b^4}
+\frac{3e_2^2}{8b^4}
-\frac{3e_2}{8b^5}.
\label{eq:M4E}
\end{equation}
Substituting these identities into the preceding expressions gives the nonlinear drag coefficients directly in terms of the membrane tension and its derivatives at the butterfly point.

The first coefficient takes the particularly simple form
\begin{equation}
c_1
=
\frac{1}{pv_B^2},
\label{eq:c1vB}
\end{equation}
while the next coefficient is
\begin{equation}
c_2
=
\frac{p+1}{2p^2v_B^4}
-
\frac{1}
{p\,v_B^5\cE''(v_B)}.
\label{eq:c2E}
\end{equation}
At the following order, we find
\begin{equation}
\begin{aligned}
c_3
=
\frac{1}{6pv_B^6}
\bigg[
&
\left(1+\frac{1}{p}\right)
\left(2+\frac{1}{p}\right)
-\frac{6\left(2+\frac{1}{p}\right)}
{v_B\cE''(v_B)}
\\
&+\frac{12}
{v_B^2\cE''(v_B)^2}
+\frac{2\cE'''(v_B)}
{v_B\cE''(v_B)^3}
\bigg].
\end{aligned}
\label{eq:c3E}
\end{equation}

These results reveal a direct correspondence between the nonlinear response of the moving quark and the local shape of the membrane near the butterfly point. The leading correction, proportional to \(v^3\), is universal within the common-frame construction: it depends only on \(v_B\), with no contribution from the membrane curvature. This follows from the endpoint conditions, which fix both the value and the first derivative of the membrane tension. The coefficient of \(v^5\) is the first to probe the curvature \(\cE''(v_B)\), while the coefficient of \(v^7\) also depends on \(\cE'''(v_B)\), describing how that curvature varies near the endpoint. More generally, the coefficient of \(v^{2n+1}\) first introduces the \(n\)-th derivative of the membrane tension.

The triangular structure also makes the reconstruction reversible. Given the coefficients of the normalized drag force, the butterfly velocity follows immediately from
\begin{equation}
v_B^2
=
\frac{1}{pc_1},
\label{eq:vBfromc1}
\end{equation}
while the second nonlinear coefficient determines the membrane curvature,
\begin{equation}
\cE''(v_B)
=
\frac{1}{
p\,v_B^5
\left[
\frac{p+1}{2}c_1^2-c_2
\right]
}.
\label{eq:E2fromc}
\end{equation}
Once these quantities are known, the next coefficient determines \(\cE'''(v_B)\), and the procedure can be continued recursively to reconstruct higher derivatives of the membrane tension. In this sense, the small-velocity drag expansion provides a local geometric reconstruction of the membrane near its butterfly endpoint.

It is important, however, to distinguish this local reconstruction from the information contained in the exact finite-velocity transformation. The coefficients \(c_n\) determine the successive derivatives of the membrane tension at a single point, whereas the complete drag curve reconstructs the membrane throughout the exterior branch. Whether the local data are sufficient to determine the full function depends on the regularity properties of the underlying geometry.

For an analytic nonextremal background, with a locally invertible radial map, the inverse-function theorem guarantees that the normalized drag function admits a convergent expansion in \(v^2\) in some neighborhood of \(v=0\). Its radius of convergence is controlled by the singularities and critical points encountered upon analytically continuing the radial relation into the complex plane. Within this analytic domain, the full collection of Taylor coefficients determines the corresponding drag function.

The situation is different when the geometry is smooth but not analytic. In that case, contributions of the form
\begin{equation}
\delta H(v^2)
\sim
\exp\left(-\frac{\mathrm{const}}{v^4}\right)
\label{eq:flatcorrection}
\end{equation}
with a positive constant vanish faster than any power as \(v\to0\). They therefore contribute to the finite-velocity force without modifying any coefficient in its Taylor expansion. Consequently, even complete knowledge of the perturbative coefficients need not uniquely determine the drag curve outside the analytic setting. The coefficient hierarchy characterizes the local horizon jet, whereas the exact tangent transformation captures the full exterior geometric dependence.

A further qualification arises for extremal or degenerate horizons, where the standard expansion in integer powers of \(v^2\) may cease to exist. Suppose that the causal function vanishes near the horizon as
\begin{equation}
a(r)
=
A_m(r-r_h)^m+\cdots,
\qquad
m>1,
\label{eq:degeneratehorizon}
\end{equation}
rather than linearly as in the nonextremal case. The worldsheet-horizon condition \(a(r_s)=v^2\) then implies
\begin{equation}
r_s-r_h
\sim
A_m^{-1/m}\lvert v\rvert^{2/m}.
\label{eq:Puiseuxrs}
\end{equation}
For a regular transverse warp factor and probe coupling near the horizon, the resulting drag force generally develops a Puiseux expansion,
\begin{equation}
F(v)
=
-\sigma v
\left[
1+\widetilde c_1\lvert v\rvert^{2/m}
+\widetilde c_2\lvert v\rvert^{4/m}
+\cdots
\right].
\label{eq:PuiseuxF}
\end{equation}
The appearance of fractional powers reflects the nonanalytic motion of the worldsheet horizon as the quark velocity approaches zero. The ordinary nonextremal coefficient hierarchy, which assumes a regular expansion in \(v^2\), therefore cannot be extended to such backgrounds by simply taking \(v_B\to0\). Nevertheless, the exact geometric transformation remains the appropriate starting point whenever the relevant membrane branch and radial relations remain well defined.

\section{Null-energy constraints on the membrane--drag correspondence}
\label{sec:NEC}

The preceding analysis established an exact correspondence between the geometric membrane tension and the stationary drag force, together with a systematic relation between the nonlinear drag coefficients and the local membrane geometry near the butterfly point. These results follow from the structure of the two variational problems and do not, by themselves, require the gravitational field equations. A natural question is therefore which membrane tensions and drag curves can actually arise from regular black-brane geometries supported by physically admissible matter.

We address this question by specializing to two-derivative Einstein gravity with matter satisfying the null energy condition (NEC). As we will show, the NEC constrains the causal-area relation \(a(Y)\) and thereby restricts the geometric membrane tension. For a constant effective string coupling, these restrictions translate directly into bounds on the complete normalized drag force and its nonlinear velocity coefficients. A particularly useful outcome is that the reduced exterior geometry can be characterized by a nonnegative function, which we will call the causal load. This function provides a unified description of how NEC matter modifies the butterfly velocity, the membrane geometry, and the stationary drag force relative to the AdS--Schwarzschild background.

\subsection{The null energy condition in membrane variables}
\label{subsec:NECgeometry}

To derive these constraints, it is convenient to work in an area gauge, in which the transverse warp factor is fixed by the radial coordinate. For an isotropic, asymptotically AdS black brane, we place the horizon at \(z=1\) and write
\begin{equation}
\dd s_E^2
=
\frac{L^2}{z^2}
\left[
-a(z)\,\dd t^2
+\frac{\dd z^2}{a(z)b(z)^2}
+\dd\bm{x}_p^2
\right],
\label{eq:areagauge}
\end{equation}
with the boundary and horizon conditions
\begin{equation}
a(0)=b(0)=1,
\qquad
a(1)=0,
\qquad
b(1)>0 .
\label{eq:areaboundary}
\end{equation}
Here \(a(z)\) is the same local causal function introduced earlier, while \(b(z)\) controls the remaining radial metric component. We consider a regular exterior branch on which \(a(z)\) decreases monotonically from the boundary to the horizon.

In this gauge, the transverse-area variable takes the particularly simple form \(Y=z^{-2p}\). The relations defining the membrane slope and its squared tension then become
\begin{equation}
Y=z^{-2p},
\qquad
q(z)
=
a(z)-\frac{z a'(z)}{2p},
\qquad
M(q(z))
=
-\frac{a'(z)}{2p\,z^{2p-1}} .
\label{eq:areamembrane}
\end{equation}
These expressions provide a direct connection between the membrane variables and the radial derivatives of the blackening function.

We now impose the NEC, which requires \(T_{MN}k^Mk^N\geq0\) for every null vector \(k^M\). For the isotropic metric \eqref{eq:areagauge}, the two independent null projections of Einstein's equations give
\begin{equation}
b'(z)
\geq
0
\label{eq:NECone}
\end{equation}
and
\begin{equation}
\frac{\dd}{\dd z}
\left(
\frac{b(z)a'(z)}{z^p}
\right)
\geq
0 .
\label{eq:NECtwo}
\end{equation}
These inequalities are the geometric manifestation of the NEC in the present ansatz and underlie the familiar holographic bounds on information velocities \cite{Mezei2017,Mezei2018}. They constrain different aspects of the matter distribution: the first is sensitive, for example, to radial scalar gradients, while the second receives contributions from electric flux and other sources of tangential null stress.

Our aim is to translate these conditions into inequalities involving the membrane tension alone. For this purpose, we introduce the dimension-dependent constants
\begin{equation}
A_p
=
\frac{p-1}{2p},
\qquad
B_p
=
\frac{p+1}{2p},
\qquad
A_p+B_p=1 .
\label{eq:ABdef}
\end{equation}
Using \eqref{eq:areamembrane}, the combination entering the second NEC inequality can be rewritten as
\begin{equation}
-\frac{b(z)a'(z)}{2p\,z^p}
=
b(Y)M(Y)Y^{-A_p},
\label{eq:NECmembraneintermediate}
\end{equation}
where \(M(Y)\) denotes the membrane tension evaluated along the radial branch parametrized by \(Y\).

Since \(Y=z^{-2p}\) decreases with increasing \(z\), the second NEC inequality implies
\begin{equation}
\frac{\dd}{\dd Y}
\left[
b(Y)M(Y)Y^{-A_p}
\right]
\geq0 .
\label{eq:NECwithb}
\end{equation}
Meanwhile, the first inequality gives \(b_Y\leq0\). Because \(b>0\) and \(M>0\) on the regular branch of interest, these two conditions can be combined to eliminate the radial function \(b\), yielding
\begin{equation}
\frac{\dd}{\dd Y}
\left[
M(Y)Y^{-A_p}
\right]
\geq0 .
\label{eq:NECmembraneY}
\end{equation}
This is an important simplification: although the original NEC constraints involve both metric functions \(a(z)\) and \(b(z)\), their combination produces a condition expressed entirely in terms of the reduced geometric data reconstructed by the membrane.

Using the Legendre relation \(Y=M'(q)\), equation \eqref{eq:NECmembraneY} becomes
\begin{equation}
M'(q)^2
\geq
A_p\,M(q)M''(q).
\label{eq:NECMq}
\end{equation}
On a regular monotone branch, the membrane also satisfies
\begin{equation}
M''(q)
=
\frac{\dd Y}{\dd q}
\geq0 .
\label{eq:Mconvex}
\end{equation}
Thus, the natural convex quantity in the membrane--drag correspondence is the squared tension \(M(q)=\cE(\sqrt q)^2\), rather than the tension \(\cE(u)\) itself. While the latter is convex on the physical interior branch under the usual assumptions, its convexity need not persist throughout the exterior continuation. The inequalities \eqref{eq:NECMq} and \eqref{eq:Mconvex} therefore provide the appropriate geometric constraints for the exterior membrane reconstruction.

These conditions also admit a converse interpretation. At the level of an effective isotropic stress tensor, they characterize the reduced geometries compatible with the Einstein--NEC equations, subject to the required regularity, monotonicity, and boundary conditions. Indeed, given a positive membrane function \(M(q)\) with \(M'>0\) satisfying the above inequalities on a regular invertible branch, one may reconstruct
\(Y=M'(q)\), \(a=q-M/M'\), and \(z=Y^{-1/(2p)}\). Choosing \(b=1\) then gives a metric whose effective stress tensor satisfies the two independent null inequalities. Additional restrictions arise when the matter content is required to follow from a specified scalar, gauge, or other microscopic Lagrangian.

The NEC therefore does more than constrain a particular radial parametrization of the metric. It restricts the class of membrane tensions, and hence the class of normalized drag curves, that can be realized by regular Einstein geometries.

\subsection{Positive causal load and exact parameterization of the NEC class}
\label{subsec:causalload}

The membrane inequality \eqref{eq:NECMq} becomes especially transparent when the local causal function itself is used as a radial variable. On the monotone exterior branch, we set
\begin{equation}
x=a .
\label{eq:xasa}
\end{equation}
This choice is particularly natural for the drag problem, since the worldsheet-horizon condition identifies \(x\) directly with the squared quark velocity, \(x=v^2\). The horizon corresponds to \(x=0\), while the asymptotic boundary is approached as \(x\to1\).

We first define the causal stiffness
\begin{equation}
\Delta(x)
=
q(x)-x .
\label{eq:Deltadef}
\end{equation}
The Legendre relations derived earlier give the equivalent expressions
\begin{equation}
\Delta
=
\frac{M}{M'}
=
Y\frac{\dd a}{\dd Y}
=
\frac{\dd a}{\dd\ln Y}.
\label{eq:Deltaequiv}
\end{equation}
The stiffness thus measures how the local causal function changes with the logarithm of the transverse-area density. It is positive on the regular exterior branch and provides a direct link between the radial variation of the geometry and the membrane slope.

In terms of this parametrization, the NEC inequality \eqref{eq:NECMq} reduces to
\begin{equation}
\frac{\dd q}{\dd x}
\geq
A_p .
\label{eq:qxbasic}
\end{equation}
This suggests introducing the nonnegative function
\begin{equation}
\cN(x)
=
\frac{\dd q}{\dd x}-A_p,
\qquad
\cN(x)\geq0 .
\label{eq:Nload}
\end{equation}
We will refer to \(\cN(x)\) as the causal load. It measures the departure of the membrane-slope response from the value that saturates the geometric NEC inequality. As we will see, this quantity also measures the departure of the reduced geometry from the AdS--Schwarzschild solution.

The usefulness of this definition is that the entire reduced exterior geometry can be reconstructed from \(\cN(x)\). The asymptotically AdS boundary condition implies \(q(1)=1\). Integrating \eqref{eq:Nload} from the boundary inward therefore gives
\begin{equation}
q(x)
=
B_p+A_px
-\int_x^1\cN(y)\,\dd y ,
\label{eq:qN}
\end{equation}
and consequently
\begin{equation}
\Delta(x)
=
B_p(1-x)
-\int_x^1\cN(y)\,\dd y .
\label{eq:DeltaN}
\end{equation}
The transverse-area density can then be reconstructed from \eqref{eq:Deltaequiv}. Since
\(\dd\ln Y/\dd x=1/\Delta(x)\), together with the horizon normalization \(Y(0)=1\), we obtain
\begin{equation}
Y(x)
=
\exp
\left[
\int_0^x
\frac{\dd y}{\Delta(y)}
\right].
\label{eq:YN}
\end{equation}
Finally, the membrane tension and slope follow parametrically as
\begin{equation}
M(x)
=
Y(x)\Delta(x),
\qquad
u(x)^2=q(x).
\label{eq:MNparam}
\end{equation}

Equations \eqref{eq:qN}--\eqref{eq:MNparam} provide an exact parameterization of the reduced exterior geometry in terms of a single nonnegative function. Subject to the positivity of \(\Delta(x)\) and the appropriate regularity and asymptotic boundary conditions, the causal load determines the relation \(a(Y)\), and therefore the exterior geometric membrane, without requiring an explicit solution for the radial lapse.

This construction also identifies AdS--Schwarzschild as a distinguished reference geometry. When \(\cN(x)=0\), the causal stiffness reduces to \(\Delta(x)=B_p(1-x)\), and the corresponding geometry is precisely the planar AdS--Schwarzschild solution. A nonzero causal load therefore provides a geometric measure of the departure from this reference background.

\subsection{Consequences for information velocities and drag}
\label{subsec:NECconsequences}

The positive causal-load representation leads directly to several bounds on the observables discussed in the preceding sections.

We begin with the butterfly velocity. At the spacetime horizon, \(x=0\), the membrane slope satisfies \(q(0)=v_B^2\). Evaluating \eqref{eq:qN} at this point gives the exact sum rule
\begin{equation}
v_B^2
=
B_p
-
\int_0^1\cN(x)\,\dd x .
\label{eq:vBsumrule}
\end{equation}
This relation provides a direct interpretation of the reduction of the butterfly velocity relative to its AdS--Schwarzschild value: the difference is precisely the integrated causal load over the exterior geometry.

Because \(\cN(x)\geq0\), the sum rule immediately implies
\begin{equation}
v_B^2
\leq
B_p
=
\frac{p+1}{2p}.
\label{eq:vBbound}
\end{equation}
The upper bound is saturated only when the causal load vanishes throughout the exterior branch. In that case, the NEC inequalities require both null channels to saturate, and the boundary-normalized geometry reduces to
\begin{equation}
b(z)=1,
\qquad
a(z)=1-z^{p+1}.
\label{eq:Schrigidity}
\end{equation}
Thus, saturation of the butterfly bound is rigid within this class: it singles out the planar AdS--Schwarzschild geometry rather than merely fixing one local horizon parameter.

The same positive profile also constrains the stationary drag force. In the common-frame case, where the effective string coupling \(Z\) is constant, the normalized force is \(H(x)=Y(x)^{1/p}\). Using \eqref{eq:YN} and \eqref{eq:DeltaN}, we obtain
\begin{equation}
H(x)
=
\exp
\left[
\frac{1}{p}
\int_0^x
\frac{\dd y}{
B_p(1-y)-\displaystyle\int_y^1\cN(s)\,\dd s
}
\right].
\label{eq:HNexact}
\end{equation}
This expression provides an exact representation of the finite-velocity drag force in terms of the causal load. In particular, it makes the effect of NEC matter transparent: a positive load reduces the causal stiffness and thereby enhances the normalized drag response.

Indeed, the positivity of \(\cN\) implies
\begin{equation}
\Delta(x)
\leq
B_p(1-x),
\label{eq:Deltabound}
\end{equation}
and therefore
\begin{equation}
H(x)
\geq
(1-x)^{-2/(p+1)} .
\label{eq:Hlowerbound}
\end{equation}
In terms of the physical quark velocity, this becomes
\begin{equation}
\lvert F(v)\rvert
\geq
\sigma v
(1-v^2)^{-2/(p+1)} .
\label{eq:Flowerbound}
\end{equation}
The right-hand side is the drag force in the neutral AdS--Schwarzschild geometry, expressed in terms of the same horizon normalization \(\sigma\).

The meaning of this comparison is important. The inequality holds at fixed horizon area and fixed common-frame string-tension normalization, rather than at fixed temperature. In a general Einstein--matter background, the relation between the temperature and entropy density may differ from that of AdS--Schwarzschild. The bound therefore constrains the normalized geometric response, not the absolute drag force at a prescribed temperature.

Moreover, the bound is rigid. If equality holds at any nonzero velocity, the nonnegative contribution of the causal load to the integral in \eqref{eq:HNexact} must vanish, forcing \(\cN=0\) throughout the exterior branch. AdS--Schwarzschild consequently provides the unique minimum-drag geometry within the regular Einstein--NEC class under the stated normalization.

The causal load can also be reconstructed directly from a given drag curve. Substituting the inverse membrane transformation \eqref{eq:qfromH} into its definition \eqref{eq:Nload}, we find
\begin{equation}
\cN(x)
=
\frac{p+3}{2p}
-
\frac{1}{p}
\frac{H(x)H''(x)}{H'(x)^2}.
\label{eq:NfromH}
\end{equation}
Thus, the NEC can be expressed entirely as a differential constraint on the normalized force,
\begin{equation}
\frac{H(x)H''(x)}{H'(x)^2}
\leq
\frac{p+3}{2}.
\label{eq:dragNECcriterion}
\end{equation}
This provides a direct consistency test for a proposed common-frame drag curve. Together with the required positivity, monotonicity, and boundary conditions, it characterizes the NEC restriction on the reduced geometry at the level of an effective isotropic stress tensor.

Expanding this inequality near the horizon yields corresponding constraints on the nonlinear drag coefficients. In particular, the butterfly bound \eqref{eq:vBbound}, combined with \(c_1=1/(pv_B^2)\), gives
\begin{equation}
c_1
\geq
\frac{2}{p+1},
\label{eq:c1NECbound}
\end{equation}
while the differential NEC condition implies
\begin{equation}
c_2
\leq
\frac{p+3}{4}c_1^2 .
\label{eq:c2NECbound}
\end{equation}
More precisely, the causal load at the horizon is
\begin{equation}
\cN(0)
=
\frac{p+3}{2p}
-
\frac{2c_2}{pc_1^2}.
\label{eq:N0coeff}
\end{equation}

These relations give the drag coefficients a direct geometric interpretation. The first nonlinear coefficient determines \(v_B\) and therefore measures the integrated causal load through \eqref{eq:vBsumrule}. Once this integrated contribution is fixed, the next coefficient determines the local value of the load at the horizon. Higher-order coefficients successively probe its derivatives near \(x=0\). The NEC thus constrains not only the overall magnitude of the drag response, but also the hierarchy of local geometric data reconstructed from its velocity expansion.

Although our emphasis is on the exterior geometry relevant to drag, the same membrane inequality also has consequences for entanglement dynamics. On the regular interior continuation, the entanglement velocity is determined by the membrane tension at the bottleneck \(q=0\),
\begin{equation}
v_E^2
=
M(0).
\label{eq:vEM0}
\end{equation}
For \(Y\leq1\), the inequality \eqref{eq:NECmembraneY} implies
\begin{equation}
M(Y)
\leq
v_B^2Y^{A_p}.
\label{eq:Minteriorbound}
\end{equation}
Using the bottleneck condition and optimizing over its radial position yields, for \(p>1\),
\begin{equation}
\frac{v_E^2}{v_B^2}
\leq
A_p^{A_p/B_p}
=
\left(
\frac{p-1}{2p}
\right)^{\frac{p-1}{p+1}},
\label{eq:vEvBbound}
\end{equation}
which, together with \eqref{eq:vBbound}, reproduces the standard AdS--Schwarzschild upper bound on the entanglement velocity.

In the present framework, these information-theoretic bounds and the drag inequalities arise from the same underlying geometric restriction. The entanglement velocity is determined by the value \(M(0)\) at the interior bottleneck, while the butterfly velocity identifies the point where the squared membrane tension meets its unit-slope tangent. The stationary drag force, in turn, probes the continuation of the same geometric structure into the exterior region. The NEC therefore provides a common set of constraints linking entanglement propagation, butterfly spreading, and momentum loss.

\subsection{Matter interpretation of the causal load}
\label{subsec:NECmatter}

So far, the causal load has been introduced entirely in geometric terms. Its physical meaning becomes more explicit when the bulk matter content is specified. As an illustrative example, consider Einstein gravity coupled to scalar fields and Abelian gauge fields,
\begin{equation}
\begin{aligned}
S
=
\frac{1}{16\pi G_N}
\int\dd^{p+2}x\sqrt{-g}
\bigg[
&R
-\frac{1}{2}\mathcal{G}_{IJ}(\phi)
\partial_M\phi^I\partial^M\phi^J
-V(\phi)
\\
&-\frac{1}{4}\mathcal{Z}_{AB}(\phi)
F^A_{MN}F^{B\,MN}
\bigg],
\end{aligned}
\label{eq:ESMaction}
\end{equation}
where the scalar and gauge kinetic matrices are positive definite. We take the scalar fields to depend only on the radial coordinate and restrict the gauge fields to purely electric configurations.

For this matter sector, the two independent null projections of Einstein's equations become
\begin{equation}
\frac{b'}{b}
=
\frac{z}{2p}
\mathcal{G}_{IJ}\phi^{I\prime}\phi^{J\prime},
\label{eq:bscalar}
\end{equation}
and
\begin{equation}
\frac{\dd}{\dd z}
\left(
\frac{ba'}{z^p}
\right)
=
z^{2-p}b\,
\mathcal{Z}_{AB}
A_t^{A\prime}A_t^{B\prime}.
\label{eq:aelectric}
\end{equation}
The positivity of the kinetic matrices makes the two NEC contributions manifest. The first is controlled by radial scalar gradients, while the second measures the contribution of the electric fields.

Combining these relations with the definition of the causal load gives
\begin{equation}
\cN(x)
=
\left.
\left[
\frac{z^2}{4p^2}
\mathcal{G}_{IJ}\phi^{I\prime}\phi^{J\prime}
+
\frac{z^3}{2p(-a')}
\mathcal{Z}_{AB}
A_t^{A\prime}A_t^{B\prime}
\right]
\right|_{a(z)=x}.
\label{eq:Nmatter}
\end{equation}
This expression makes the positivity of \(\cN(x)\) explicit. Scalar gradients and electric flux contribute additively to the causal load, with each term nonnegative on the regular exterior branch. The scalar potential does not appear explicitly because its contribution to the stress tensor is proportional to the metric and therefore vanishes upon contraction with a null vector.

The causal load can thus be interpreted as a geometrically weighted measure of the matter contributions that modify the local causal structure of the black brane. Its integral controls the reduction of the butterfly velocity, while its local value and derivatives determine the successive corrections to the normalized drag force. In this sense, the same function provides a direct bridge between the bulk matter distribution, the geometric membrane, and the nonlinear response of a moving quark.

It is important to emphasize that the resulting drag bounds apply to the common-frame construction, in which the effective string coupling is constant. The geometric NEC constraints themselves remain valid independently of the choice of string probe. When the effective string tension depends on the radial position, however, the force acquires additional probe-specific contributions that are not determined by the Einstein-frame membrane alone. We turn to this distinction in the following section.

\section{Probe-frame dressing and the limits of membrane universality}
\label{sec:dressing}

The preceding sections established that the geometric membrane determines the causal-area relation \(a(Y)\), which in turn fixes the stationary drag force when the effective string coupling is constant. This correspondence, however, relies on an important distinction between the background geometry and the coupling of the string probe to that geometry. While the membrane tension is constructed from the Einstein-frame metric, the drag force is governed by the effective string tension, which may acquire additional radial dependence through scalar fields or the transformation between the Einstein and string frames. The membrane therefore reconstructs the geometric contribution to the drag force, but does not necessarily determine its complete dependence on the quark velocity.

This distinction is absent in conformal backgrounds with a constant dilaton, where the Einstein-frame and string-frame metrics differ only by a constant factor. It becomes essential in more general settings, particularly Einstein--dilaton backgrounds and bottom-up holographic QCD models, where the running scalar fields modify the effective string tension along the radial direction. Our aim in this section is to identify precisely how this additional dependence enters the membrane--drag transformation and to establish which aspects of the force remain determined by the underlying geometry.

To separate the geometric and probe-dependent contributions, we introduce the dimensionless coupling
\begin{equation}
\omega(Y)
=
\frac{Z(Y)}{Z_h},
\qquad
\omega(1)=1 .
\label{eq:omegadef}
\end{equation}
Here \(Z(Y)\) is the effective string coupling expressed in terms of the transverse-area variable, while \(Z_h\) denotes its value at the spacetime horizon. The normalization ensures that the additional dressing disappears in the zero-velocity limit.

Combining the general trailing-string result \eqref{eq:HZradial} with the membrane transformation, the normalized drag force becomes
\begin{equation}
H_Z(x)
=
\omega(Y_x)Y_x^{1/p},
\qquad
Y_x
=
\min_{q>x}
\frac{M(q)}{q-x}.
\label{eq:dressedtransform}
\end{equation}
The minimum is understood on the regular convex exterior branch, where the tangent construction selects the worldsheet-horizon geometry. This expression shows that the membrane continues to determine the transverse-area density \(Y_x\) associated with a given quark velocity. The additional factor \(\omega(Y_x)\), however, describes how the chosen string probe couples to that geometry.

The complete classical force is therefore specified by three ingredients,
\begin{equation}
\left\{
M(q),\,
\omega(Y),\,
\sigma
\right\}.
\label{eq:minimaldata}
\end{equation}
The membrane tension \(M(q)\) determines the reduced causal-area geometry, the function \(\omega(Y)\) encodes the radial variation of the probe coupling, and \(\sigma=T_fZ_hC_h\) fixes the overall force normalization. In the common-frame limit, \(\omega(Y)=1\), the normalized force reduces to the purely geometric membrane transformation derived earlier.

The distinction between these contributions becomes particularly transparent in the small-velocity expansion. To characterize the radial dependence of the probe coupling near the horizon, we define
\begin{equation}
\omega_j
=
\left.
\frac{\dd^j\omega}{\dd Y^j}
\right|_{Y=1}.
\label{eq:omegaj}
\end{equation}
Expanding the dressed force around \(x=v^2=0\), the coefficient of the leading nonlinear correction is
\begin{equation}
c_1^{(Z)}
=
\frac{1/p+\omega_1}{v_B^2}.
\label{eq:c1dressed}
\end{equation}
Equivalently, in an arbitrary radial coordinate, this relation can be written as
\begin{equation}
c_1^{(Z)}
=
\frac{1}{pv_B^2}
+
\frac{(\log Z)'_h}{a'_h}.
\label{eq:c1dressedradial}
\end{equation}
The first term is the geometric contribution previously identified with the butterfly velocity, while the second depends explicitly on the radial variation of the effective string coupling at the horizon. Consequently, the butterfly velocity alone does not determine the leading nonlinear drag coefficient for a general string probe.

This correction has a direct physical realization in improved holographic QCD. In such models, the Einstein-frame and string-frame metrics are related by a dilaton-dependent Weyl transformation, and the drag force must be computed using the string-frame geometry \cite{GursoyEtAl2009}. The running dilaton then contributes an additional horizon-gradient term to the cubic drag coefficient. The geometric relation derived in the previous sections remains intact, but the complete force contains information about the scalar coupling that is not encoded in the Einstein-frame membrane tension.

The necessity of this additional information can be demonstrated without specifying a particular matter model. Consider a fixed Einstein-frame background and a reference effective string coupling \(Z_0(r)\). We introduce a family of probe couplings
\begin{equation}
Z_{\lambda,N}(r)
=
Z_0(r)
\left[
1+\lambda a(r)^N
\right],
\qquad
N\geq1,
\label{eq:noGofamily}
\end{equation}
where \(\lambda\) is chosen such that the effective string coupling remains positive throughout the exterior geometry.

Since \(a(r_h)=0\), all members of this family have the same horizon coupling, \(Z_{\lambda,N}(r_h)=Z_0(r_h)\), and therefore the same force normalization \(\sigma\). At the worldsheet horizon, however, the defining condition \(a(r_s)=v^2\) gives
\begin{equation}
F_{\lambda,N}(v)
=
F_0(v)
\left[
1+\lambda v^{2N}
\right].
\label{eq:noGoforce}
\end{equation}
Thus, varying \(\lambda\) changes the velocity dependence of the drag force without modifying the background geometry.

In particular, the complete Einstein-frame membrane tension, the thermal entropy density, and the information velocities \(v_B\) and \(v_E\) remain unchanged. The nonlinear drag coefficients below order \(v^{2N+1}\) are also unaffected, whereas the coefficient at that order acquires an additional contribution proportional to \(\lambda\). Since \(N\) can be chosen arbitrarily, this construction shows that probe-dependent contributions can first appear at any prescribed order in the velocity expansion while leaving all lower-order coefficients and the entire geometric membrane unchanged.

This establishes an important limitation of membrane universality. The geometric membrane determines the drag force only after the coupling of the string probe has been specified. In the absence of that information, neither the exact membrane tension nor its complete set of endpoint derivatives is sufficient to reconstruct the nonlinear force coefficients of an arbitrary probe sector. In particular, the geometric bounds on the common-frame drag force derived from the NEC do not automatically extend to a radially dressed string coupling.

The same distinction, however, also suggests a complementary reconstruction procedure. If both the Einstein-frame membrane tension and the dressed drag force are known, the membrane transformation first determines the transverse-area variable \(Y(x)\) at the worldsheet horizon. The normalized probe coupling can then be extracted directly from
\begin{equation}
\omega(Y(x))
=
\frac{H_Z(x)}{Y(x)^{1/p}}.
\label{eq:dilatontomography}
\end{equation}
Thus, independent knowledge of the entanglement membrane and the stationary drag force allows the exterior probe coupling to be reconstructed as a function of the transverse-area density.

The membrane--drag correspondence therefore provides a natural separation between two types of bulk information. The geometric membrane determines the causal-area structure of the thermal medium, while the difference between the measured drag force and its purely geometric prediction isolates the additional coupling carried by the string probe. Rather than invalidating the correspondence, probe-frame dressing identifies its precise domain of universality and extends its reconstruction framework to include defect-specific information.

\section{Some explicit examples}
\label{sec:examples}

Having established the exact membrane--drag correspondence and the constraints imposed by the null energy condition, we now illustrate these results in explicit black-brane geometries. We focus on backgrounds for which the relation between the local causal function and the transverse-area density can be determined analytically, allowing the membrane tension and drag force to be obtained without relying on a small-velocity expansion.

The examples serve complementary purposes. The power-law family provides a simple setting in which the complete velocity dependence of the drag force is determined by the butterfly velocity alone. Its AdS--Schwarzschild limit gives an explicit realization of the geometry saturating the NEC bounds derived above. Charged AdS--Reissner--Nordstr\"om backgrounds, by contrast, provide a nontrivial test of the correspondence in the presence of matter, while their extremal limit illustrates the emergence of nonanalytic corrections beyond the ordinary velocity expansion. Throughout this section, we consider the common-frame string probe with constant effective coupling \(Z\).

\subsection*{Planar AdS--Schwarzschild and the power-law family}
\label{subsec:schwarzschild}

We begin with a class of geometries characterized by power-law radial dependence,
\begin{equation}
\frac{C(y)}{C_h}
=
y^{-2\alpha},
\qquad
a(y)
=
1-y^m,
\qquad
y_h=1 .
\label{eq:powerlawgeometry}
\end{equation}
Here \(\alpha\) and \(m\) are positive parameters, and the radial coordinate is chosen so that the horizon lies at \(y=1\) and the asymptotic boundary is approached as \(y\to0\). This family provides a useful analytic realization of the membrane--drag transformation, since the causal-area relation takes a simple power-law form.

The normalized transverse-area density is
\(Y=y^{-2\alpha p}\). Evaluating the membrane-slope relation at the horizon gives the butterfly velocity,
\begin{equation}
v_B^2
=
\frac{m}{2\alpha p}.
\label{eq:powerlawvB}
\end{equation}
The worldsheet horizon is determined by \(a(y_s)=x\), with \(x=v^2\), which immediately yields \(y_s^m=1-x\). The normalized drag force therefore takes the exact form
\begin{equation}
H(x)
=
(1-x)^{-2\alpha/m}
=
(1-x)^{-1/(pv_B^2)}.
\label{eq:powerlawH}
\end{equation}
The corresponding squared membrane tension, obtained by eliminating the radial coordinate through the Legendre relations, is
\begin{equation}
M(q)
=
v_B^2
\left[
\frac{1-v_B^2}{1-q}
\right]^{(1-v_B^2)/v_B^2}.
\label{eq:powerlawM}
\end{equation}
Here we consider the regular invertible branch with \(0<v_B^2<1\).

These expressions provide an explicit realization of the exact tangent transformation. More importantly, they reveal a special property of the power-law family: the complete normalized drag curve is determined by the single parameter \(v_B\). This occurs because the causal-area relation is itself a pure power law, leaving no additional independent geometric information beyond its exponent. Consequently, all nonlinear drag coefficients are fixed once the butterfly velocity is known.

This property is not generic. For a general black-brane geometry, the butterfly velocity determines only the leading nonlinear coefficient, while higher orders probe additional derivatives of the membrane tension at the butterfly point. The power-law family therefore represents a particularly simple case in which the local near-horizon information is sufficient to determine the entire finite-velocity response.

An important member of this family is the planar AdS--Schwarzschild black brane, for which
\begin{equation}
m=p+1,
\qquad
\alpha=1,
\qquad
v_B^2=B_p.
\label{eq:Schparameters}
\end{equation}
The normalized drag force becomes
\begin{equation}
H_{\mathrm{Sch}}(x)
=
(1-x)^{-2/(p+1)},
\label{eq:SchH}
\end{equation}
and its expansion coefficients are given exactly by
\begin{equation}
c_n^{\mathrm{Sch}}
=
\frac{
\left(\frac{2}{p+1}\right)_n
}{n!}.
\label{eq:Schcoefficients}
\end{equation}
Here \((a)_n\) denotes the rising Pochhammer symbol. These coefficients provide an all-order analytic benchmark for the general membrane expansion derived in \eqref{eq:allordercn}.

The AdS--Schwarzschild solution also has a distinguished role in the NEC analysis. Its causal load vanishes identically, \(\cN(x)=0\), so that the geometry saturates the butterfly-velocity bound, the entanglement-velocity bound, and the minimum-drag inequality. The differential NEC constraint is likewise saturated throughout the exterior branch. Thus, the same solution provides both an exact realization of the membrane--drag transformation and the reference geometry against which the effects of NEC matter can be measured.

For \(p=3\), corresponding to a four-dimensional boundary field theory, the normalized drag force reduces to the familiar expression
\begin{equation}
H_{\mathrm{Sch}}(v^2)
=
\frac{1}{\sqrt{1-v^2}},
\label{eq:AdS5H}
\end{equation}
with expansion coefficients
\begin{equation}
c_1=\frac{1}{2},
\qquad
c_2=\frac{3}{8},
\qquad
c_3=\frac{5}{16},
\qquad
c_4=\frac{35}{128}.
\label{eq:AdS5coefficients}
\end{equation}
These values agree with the general coefficient hierarchy and provide a direct consistency check of the membrane endpoint relations.

\subsection*{Charged AdS--Reissner--Nordstr\"om}
\label{subsec:RN}

We next consider a charged black brane, which provides a more general example in which the matter content modifies the causal-area geometry. Unlike the power-law family, the finite-velocity drag response is no longer determined solely by the butterfly velocity through a universal power-law dependence. This makes charged AdS--Reissner--Nordstr\"om an instructive setting for testing both the exact membrane transformation and the NEC constraints.

Working in the area gauge introduced in \eqref{eq:areagauge}, with the horizon located at \(z=1\), the blackening function for the planar AdS--Reissner--Nordstr\"om geometry in \(p>1\) spatial boundary dimensions is
\begin{equation}
a(z)
=
1-(1+Q^2)z^{p+1}
+Q^2z^{2p},
\qquad
b(z)=1 .
\label{eq:RNblackening}
\end{equation}
The parameter \(Q\) controls the charge of the black brane, and the neutral limit \(Q=0\) reproduces AdS--Schwarzschild.

Evaluating the membrane-slope relation at the horizon gives
\begin{equation}
v_B^2
=
B_p-A_pQ^2,
\label{eq:RNvB}
\end{equation}
showing explicitly how the charge reduces the butterfly velocity. The extremal limit is reached at
\(Q^2=(p+1)/(p-1)\), where the temperature and butterfly velocity vanish.

For a constant effective string coupling, the normalized drag force is determined by the transverse warp factor at the worldsheet horizon, \(H(x)=z_s^{-2}\). Substituting this relation into the worldsheet-horizon condition \(a(z_s)=x\), we obtain
\begin{equation}
x
=
1-(1+Q^2)H(x)^{-(p+1)/2}
+Q^2H(x)^{-p}.
\label{eq:RNH}
\end{equation}
This algebraic equation determines the exact finite-velocity drag curve, even when an explicit closed-form solution for \(H(x)\) is not available.

The corresponding geometric membrane can also be obtained analytically. Using the radial Legendre relations, we find
\begin{equation}
M(q)
=
B_p(1+Q^2)
\left[
\frac{A_p(1+Q^2)}
{1-q}
\right]^{A_p/B_p}
-Q^2 .
\label{eq:RNM}
\end{equation}
Substituting this expression into the tangent transformation reproduces \eqref{eq:RNH}, providing a nontrivial verification of the exact membrane--drag correspondence in a charged background.

The effect of charge is also consistent with the general NEC analysis. For \(Q\neq0\), the electric field generates a positive causal load, which reduces the butterfly velocity relative to its Schwarzschild value. At the same time, the normalized drag force is enhanced and lies strictly above the Schwarzschild curve for every nonzero subluminal velocity. The charged solution therefore illustrates how matter contributions modify the causal-area geometry while preserving the exact relation between the membrane tension and the common-frame drag force.

The extremal limit reveals a further feature of the correspondence. As discussed in the general analysis, the ordinary expansion in integer powers of \(v^2\) relies on a nondegenerate horizon. When the horizon becomes extremal, the butterfly endpoint collapses to \(v_B=0\), and the relation between the worldsheet-horizon position and the quark velocity becomes nonanalytic.

To see this explicitly, consider extremal AdS\(_5\)--Reissner--Nordstr\"om, corresponding to \(p=3\) and \(Q^2=2\). In this case, the exact drag relation reduces to
\begin{equation}
x
=
1-3H^{-2}+2H^{-3}.
\label{eq:RNextremalH}
\end{equation}
Expanding the physical solution near \(v=0\), one finds
\begin{equation}
H(v^2)
=
1+\frac{\lvert v\rvert}{\sqrt{3}}
+\frac{4}{9}v^2+\cdots,
\label{eq:RNextremalseries}
\end{equation}
and hence
\begin{equation}
F(v)
=
-\sigma v
\left[
1+\frac{\lvert v\rvert}{\sqrt{3}}
+\frac{4}{9}v^2+\cdots
\right].
\label{eq:RNextremalforce}
\end{equation}

The appearance of the term \(v\lvert v\rvert\) demonstrates the breakdown of the ordinary odd-power expansion characteristic of nonextremal horizons. Rather than appearing as a regular series in \(v^2\), the normalized force develops a contribution proportional to \(\lvert v\rvert\), reflecting the degenerate behavior of the causal function near the extremal horizon.

This result illustrates an important distinction between the exact geometric correspondence and its perturbative expansion. The finite-velocity membrane--drag relation continues to hold in the extremal background, but the local Taylor hierarchy around \(v=0\) is replaced by a nonanalytic expansion. The extremal solution therefore provides a concrete example of the broader validity of the exact transformation, as well as the limitations of the nonextremal coefficient formulas.

\section{Scalar pole-skipping as a reconstruction of drag and membrane data}
\label{sec:poleskipping}

The preceding analysis established that the nonlinear velocity expansion of the drag force encodes the near-horizon geometry, with successive coefficients probing higher derivatives of the membrane tension at the butterfly point. This suggests a natural question: whether the same geometric information can be reconstructed independently from the spectral properties of boundary correlation functions. We now show that such a reconstruction is possible through the lower-half-plane scalar pole-skipping tower, which provides a complementary characterization of the near-horizon data governing both the drag force and the geometric membrane.

In holography, pole-skipping occurs at special complex frequencies and momenta where the near-horizon equations for linearized bulk perturbations become degenerate. For a minimally coupled scalar, these degeneracies are determined by the successive coefficients in the near-horizon expansion of the background metric. As a result, the skipped momenta provide an independent set of spectral data from which the local bulk geometry can be reconstructed. Combined with the membrane--drag correspondence developed above, this establishes a systematic connection between scalar pole-skipping, nonlinear momentum loss, and the local shape of the entanglement membrane.

The relation is most transparent in area-radius gauge, where the transverse spatial metric is fixed to \(r^2\delta_{ij}\). We write the black-brane geometry in ingoing Eddington--Finkelstein coordinates as
\begin{equation}
\dd s^2
=
-f(r)\,\dd v^2
+2\beta(r)\,\dd v\,\dd r
+r^2\dd\bm{x}_p^2,
\label{eq:PSmetric}
\end{equation}
and place the horizon at \(r_h=1\). For a regular, nonextremal horizon, the metric functions admit expansions
\begin{equation}
f(r)
=
\sum_{j=1}^{\infty}f_j(r-1)^j,
\qquad
\beta(r)
=
\sum_{j=0}^{\infty}\beta_j(r-1)^j .
\label{eq:PSmetricexpansion}
\end{equation}
The leading horizon coefficients determine the surface gravity and butterfly velocity,
\begin{equation}
\kappa
=
2\pi T
=
\frac{f_1}{2\beta_0},
\qquad
v_B^2
=
\frac{f_1}{2p}.
\label{eq:PSkappavB}
\end{equation}

For a common-frame string probe, the worldsheet horizon is determined by
\begin{equation}
x=v^2
=
\frac{f(r_s)}{r_s^2},
\label{eq:PSworldsheet}
\end{equation}
while the normalized drag force takes the particularly simple form
\begin{equation}
H(x)
=
r_s(x)^2 .
\label{eq:PSH}
\end{equation}
Thus, the complete velocity dependence of the stationary drag force is encoded in the radial relation between \(f(r)\) and the position of the worldsheet horizon.

This observation allows the nonlinear drag coefficients to be expressed directly in terms of the near-horizon expansion of the blackening function. Introducing \(\rho=r_s-1\), the worldsheet-horizon condition becomes \(x=f(1+\rho)/(1+\rho)^2\). Applying the Lagrange--B\"urmann inversion formula gives
\begin{equation}
c_n
=
\frac{2}{n}
[\rho^{\,n-1}]
(1+\rho)^{2n+1}
\left[
\frac{\rho}{f(1+\rho)}
\right]^n .
\label{eq:cnfjallorder}
\end{equation}
Here, the bracket notation denotes extraction of the corresponding power-series coefficient. An immediate consequence is that \(c_n\) depends only on the first \(n\) coefficients \(f_1,\ldots,f_n\) of the near-horizon metric expansion. In particular,
\begin{equation}
c_1
=
\frac{2}{f_1},
\label{eq:c1f}
\end{equation}
\begin{equation}
c_2
=
\frac{5f_1-2f_2}{f_1^3},
\label{eq:c2f}
\end{equation}
and
\begin{equation}
c_3
=
\frac{
14f_1^2-14f_1f_2+4f_2^2-2f_1f_3
}{
f_1^5
}.
\label{eq:c3f}
\end{equation}
These expressions provide a direct geometric interpretation of the nonlinear velocity expansion: each successive drag coefficient resolves one additional derivative of the blackening function at the horizon.

Importantly, the radial metric function \(\beta(r)\) does not enter these relations. This is the same insensitivity to the radial lapse encountered in the exact membrane--drag correspondence. Scalar wave propagation, however, depends on both \(f(r)\) and \(\beta(r)\). The pole-skipping spectrum can therefore encode additional near-horizon geometric information that is inaccessible to the stationary drag force or the geometric membrane alone.

To make this connection explicit, consider a minimally coupled scalar field of mass \(m\), described by
\begin{equation}
S_\Phi
=
-\frac{1}{2}
\int\dd^{p+2}x\sqrt{-g}
\left[
g^{MN}\partial_M\Phi\partial_N\Phi
+m^2\Phi^2
\right].
\label{eq:scalaraction}
\end{equation}
Using the Fourier decomposition
\begin{equation}
\Phi(v,r,\bm{x})
=
e^{-i\omega v+i kx}\phi(r),
\qquad
\mu=k^2,
\label{eq:scalarFourier}
\end{equation}
the scalar equation of motion becomes
\begin{equation}
\begin{aligned}
0={}&
\frac{f}{\beta^2}\phi''
+
\left[
-\frac{2i\omega}{\beta}
+\frac{f'}{\beta^2}
+\frac{pf}{r\beta^2}
-\frac{f\beta'}{\beta^3}
\right]\phi'
\\
&-
\left[
\frac{\mu}{r^2}
+\frac{ip\omega}{r\beta}
+m^2
\right]\phi .
\end{aligned}
\label{eq:scalarODE}
\end{equation}

For generic values of \((\omega,\mu)\), regularity at the future horizon selects a unique ingoing solution. At the negative imaginary Matsubara frequencies
\begin{equation}
\omega_n
=
-in\kappa,
\qquad
n=1,2,\ldots,
\label{eq:Matsubarafrequencies}
\end{equation}
the structure of the near-horizon recursion changes. At special values of \(\mu\), the recursion relations become degenerate, allowing a second regular ingoing solution and generically producing an ambiguity in the corresponding retarded Green function. This is the scalar pole-skipping phenomenon discussed in \cite{BlakeDavisonVegh2020}.

At each level \(n\), the degeneracy condition is expressed as the vanishing of an \(n\times n\) near-horizon determinant, which defines a polynomial of degree \(n\) in \(\mu\). Its roots provide the skipped momenta at that level. Because the polynomial coefficients depend on the near-horizon expansion of the metric, these momenta supply a discrete spectral characterization of the local bulk geometry. We now show how this information determines the nonlinear drag coefficients and the membrane endpoint data.

\subsection{The first two pole-skipping levels}
\label{subsec:firsttwoPS}

The first pole-skipping level already establishes a direct connection between the scalar spectrum, the butterfly velocity, and the leading nonlinear correction to the drag force.

At the frequency \(\omega_1=-i\kappa\), the coefficient multiplying the first radial derivative in the leading horizon equation vanishes. The remaining condition becomes
\begin{equation}
\left[
\mu_{1,1}+m^2+\frac{p\kappa}{\beta_0}
\right]\phi_0
=
0 .
\label{eq:firstPSequation}
\end{equation}
The first skipped momentum therefore satisfies
\begin{equation}
\mu_{1,1}+m^2
=
-\frac{p\kappa}{\beta_0}
=
-\frac{\kappa^2}{v_B^2}.
\label{eq:firstPSvB}
\end{equation}
This relation shows that the first scalar pole-skipping point is governed by the same near-horizon geometric scale that determines the butterfly velocity.

For a massless scalar, we introduce the notation
\begin{equation}
\epsilon
=
\mu_{1,1}.
\label{eq:epsilondef}
\end{equation}
The butterfly velocity can then be reconstructed directly from the skipped momentum,
\begin{equation}
v_B^2
=
-\frac{\kappa^2}{\epsilon},
\label{eq:vBepsilon}
\end{equation}
and the geometric relation \(c_1=1/(pv_B^2)\) immediately yields
\begin{equation}
c_1
=
-\frac{\epsilon}{p\kappa^2}.
\label{eq:c1epsilon}
\end{equation}
Thus, the first nonlinear drag coefficient is determined entirely by the first massless scalar pole-skipping momentum and the temperature. This provides a direct spectral interpretation of the leading correction to the stationary momentum loss of a heavy quark.

There is also a connection with the pole-skipping phenomenon associated with quantum chaos. The first lower-half-plane pole-skipping point of a minimally coupled massless scalar has the same complex momentum as the upper-half-plane energy-density pole-skipping point,
\begin{equation}
k_\chi^2
=
-\frac{\kappa^2}{v_B^2},
\qquad
\omega_\chi=+i\kappa,
\label{eq:chaosPS}
\end{equation}
whose relation to shockwave propagation and the butterfly velocity is universal in Einstein gravity coupled to matter \cite{BlakeDavisonGrozdanovLiu2018}.

This equality of momenta should not be interpreted as identifying generic scalar pole-skipping with the chaotic energy-density mode. Rather, it reflects their common dependence on the transverse near-horizon geometry. The first scalar degeneracy and the chaos-related energy-density degeneracy occur at different frequencies, but their momenta are fixed by the same butterfly scale. It is this shared geometric dependence that allows the first scalar pole-skipping point to determine the leading nonlinear drag coefficient.

The second pole-skipping level probes additional information about the radial structure of the background. At the frequency \(\omega_2=-2i\kappa\), the degeneracy condition produces two skipped momenta,
\begin{equation}
\mu_{2,1},
\qquad
\mu_{2,2}.
\label{eq:secondroots}
\end{equation}
It is convenient to organize these roots through their elementary symmetric combinations,
\begin{equation}
S_2
=
\mu_{2,1}+\mu_{2,2},
\qquad
P_2
=
\mu_{2,1}\mu_{2,2}.
\label{eq:S2P2}
\end{equation}

The first pole-skipping level already determines the leading near-horizon metric coefficients,
\begin{equation}
f_1
=
-\frac{2p\kappa^2}{\epsilon},
\qquad
\beta_0
=
-\frac{p\kappa}{\epsilon}.
\label{eq:f1beta0PS}
\end{equation}
Expanding the scalar equation to the next order near the horizon and matching the coefficients of the resulting quadratic pole-skipping polynomial, we obtain
\begin{equation}
f_2
=
-\kappa^2
\left[
\frac{p^2P_2}{4\epsilon^3}
+\frac{2p^2}{\epsilon}
-\frac{p^2S_2}{\epsilon^2}
+\frac{3p}{\epsilon}
\right],
\label{eq:f2PS}
\end{equation}
and
\begin{equation}
\beta_1
=
\kappa
\left[
\frac{p^2S_2}{2\epsilon^2}
-\frac{p^2}{\epsilon}
-\frac{p^2P_2}{4\epsilon^3}
-\frac{p}{\epsilon}
\right].
\label{eq:beta1PS}
\end{equation}
The two independent combinations of second-level momenta therefore determine both the next coefficient of the blackening function and the first radial correction to the lapse function.

Only the former is required for the nonlinear drag force. Substituting the reconstructed value of \(f_2\) into \eqref{eq:c2f}, we find
\begin{equation}
c_2
=
\frac{
4p\epsilon S_2
-pP_2
-8(p-1)\epsilon^2
}{
16p^2\kappa^4
}.
\label{eq:c2PS}
\end{equation}
Together with the first-level result \eqref{eq:c1epsilon}, this determines the drag force through order \(v^5\),
\begin{equation}
\begin{aligned}
F(v)
=
-\sigma\bigg[
v
&-\frac{\epsilon}{p\kappa^2}v^3
\\
&+
\frac{
4p\epsilon S_2
-pP_2
-8(p-1)\epsilon^2
}{
16p^2\kappa^4
}
v^5
+O(v^7)
\bigg].
\end{aligned}
\label{eq:Fthroughv5PS}
\end{equation}
The first two scalar pole-skipping levels thus provide a complete spectral determination of the leading two nonlinear drag coefficients in the common-frame construction.

The same information also reconstructs the local shape of the geometric membrane near the butterfly point. Recall that the coefficient \(c_2\) is sensitive to the curvature of the membrane tension at \(u=v_B\). Differentiating the inverse membrane transformation and expressing the result in terms of the pole-skipping data gives
\begin{equation}
\left.
\frac{\dd q}{\dd x}
\right|_{x=0}
=
\frac{
P_2-4\epsilon S_2+16\epsilon^2
}{
8\epsilon^2}.
\label{eq:qprimePS}
\end{equation}
On the other hand, the membrane endpoint relations imply
\begin{equation}
\left.
\frac{\dd q}{\dd x}
\right|_{x=0}
=
\frac{2}{v_B\cE''(v_B)}.
\label{eq:qprimeE}
\end{equation}
Combining these expressions yields
\begin{equation}
\cE''(v_B)
=
\frac{
16\epsilon^2
}{
v_B
\left[
P_2-4\epsilon S_2+16\epsilon^2
\right]
}.
\label{eq:E2PS}
\end{equation}

We therefore obtain a direct geometric interpretation of the first two pole-skipping levels. The first skipped momentum determines the position of the butterfly point on the membrane curve, while the two second-level momenta determine the curvature of the membrane tension at that point. The same spectral data simultaneously fix the \(v^3\) and \(v^5\) corrections to the stationary drag force.

Although the preceding expressions were written for a massless scalar, the reconstruction can be extended to a scalar with constant nonzero mass. It is convenient to introduce the shifted momenta
\begin{equation}
\bar\epsilon
=
\mu_{1,1}+m^2,
\qquad
\bar\mu_{2,a}
=
\mu_{2,a}+m^2,
\label{eq:shiftedroots}
\end{equation}
together with their symmetric combinations \(\bar S_2\) and \(\bar P_2\). The corresponding drag coefficients are
\begin{equation}
c_1
=
-\frac{\bar\epsilon}{p\kappa^2},
\label{eq:c1massivePS}
\end{equation}
and
\begin{equation}
c_2
=
\frac{
4p\bar\epsilon\bar S_2
-p\bar P_2
+4m^2\bar\epsilon
-8(p-1)\bar\epsilon^2
}{
16p^2\kappa^4
}.
\label{eq:c2massivePS}
\end{equation}
The residual mass-dependent contribution in the second coefficient reflects the different radial dependence of the momentum and mass terms in the scalar equation. While spatial momentum enters through \(k^2/r^2\), a constant bulk mass has no corresponding radial factor. Consequently, the mass dependence cannot be removed entirely by a uniform shift of the skipped momenta.

These results illustrate the basic mechanism of the spectral reconstruction: the pole-skipping momenta determine successive coefficients of the near-horizon metric, which in turn determine the nonlinear drag response and the local membrane geometry. We now show how this correspondence extends systematically to arbitrary pole-skipping order.

\subsection{All-order triangular reconstruction}
\label{subsec:allorderPS}

The structure revealed by the first two pole-skipping levels extends to the entire lower-half-plane tower. At each successive Matsubara frequency, the scalar horizon equations probe higher derivatives of the background metric. These additional geometric data determine the next nonlinear drag coefficient and, through the membrane--drag transformation, the corresponding higher derivative of the membrane tension at the butterfly point.

To formulate this reconstruction, we expand the radial scalar profile near the horizon as
\begin{equation}
\phi(r)
=
\sum_{j=0}^{\infty}
\phi_j(r-1)^j .
\label{eq:phihorizonseries}
\end{equation}
Substitution into the radial equation produces a hierarchy of algebraic recursion relations for the coefficients \(\phi_j\). At the special frequency \(\omega_n=-in\kappa\), the first \(n\) recursion equations can be organized into the matrix system
\begin{equation}
\cM^{(n)}(\mu)\,
\begin{pmatrix}
\phi_0\\
\phi_1\\
\vdots\\
\phi_{n-1}
\end{pmatrix}
=
0 .
\label{eq:recursionmatrix}
\end{equation}
The corresponding pole-skipping momenta are determined by the degeneracy condition
\begin{equation}
\det\cM^{(n)}(\mu)
=
0 .
\label{eq:PSdet}
\end{equation}
For a minimally coupled scalar, this determinant defines a polynomial of degree \(n\) in \(\mu=k^2\). Denoting its roots by \(\mu_{n,a}\), with \(a=1,\ldots,n\), we introduce their elementary symmetric combinations
\begin{equation}
e_{n,j}
=
\sum_{1\leq a_1<\cdots<a_j\leq n}
\mu_{n,a_1}\cdots\mu_{n,a_j}.
\label{eq:enjdef}
\end{equation}
These quantities characterize the pole-skipping polynomial independently of the ordering of its roots and provide the natural variables for reconstructing the near-horizon metric.

The reconstruction proceeds recursively. Once the lower-order metric coefficients have been determined from the preceding pole-skipping levels, the new information at level \(n\geq2\) is contained in the two highest-degree symmetric combinations,
\begin{equation}
e_{n,n-1},
\qquad
e_{n,n}.
\label{eq:twoindependentPS}
\end{equation}
These determine the two new coefficients in the near-horizon metric expansion,
\begin{equation}
f_n,
\qquad
\beta_{n-1}.
\label{eq:twometriccoeffs}
\end{equation}
The remaining \(n-2\) symmetric combinations do not introduce additional independent geometric data. Instead, they satisfy polynomial consistency relations involving coefficients already reconstructed at lower orders. This triangular reconstruction, including its redundancy structure and extension to arbitrary order, was established in \cite{LuRanWu2026}.

Combining this result with the all-order drag formula \eqref{eq:cnfjallorder} gives a direct map from the scalar pole-skipping spectrum to the nonlinear drag coefficients,
\begin{equation}
\left\{
\mu_{m,a};
\ 1\leq m\leq n
\right\}
\longrightarrow
\left\{
f_j,\beta_{j-1};
\ 1\leq j\leq n
\right\}
\longrightarrow
c_n .
\label{eq:PStodragchain}
\end{equation}
The significance of this relation is that the pole-skipping data reconstruct the full near-horizon metric information required at each order, whereas the stationary drag force depends only on a subset of that information.

More precisely, once the lower-order coefficients are known, the new contribution to \(c_n\) is determined by \(f_n\). The simultaneously reconstructed coefficient \(\beta_{n-1}\) does not enter the mean drag force, reflecting its insensitivity to the radial lapse. Thus, each pole-skipping level contains one new piece of geometric information shared with the membrane and drag force, together with an additional radial-lapse datum that these observables cannot determine. The remaining spectral combinations impose consistency conditions on the reconstructed geometry.

The third nonlinear drag coefficient provides the first illustration beyond the explicit second-level analysis. Once the level-three pole-skipping data have been used to reconstruct \(f_3\), the next coefficient follows from
\begin{equation}
c_3
=
\frac{
14f_1^2-14f_1f_2+4f_2^2-2f_1f_3
}{
f_1^5}.
\label{eq:c3PSalgorithm}
\end{equation}
Through the membrane relation \eqref{eq:c3E}, the same information determines \(\cE'''(v_B)\), once the lower-order endpoint data have been fixed. More generally, the scalar pole-skipping tower through level \(n\) reconstructs the drag force through order \(v^{2n+1}\) and the membrane endpoint derivatives through \(\cE^{(n)}(v_B)\).

This establishes an all-order correspondence between two seemingly different types of boundary information. The nonlinear drag coefficients characterize the response of a heavy external quark moving through the thermal medium, while the pole-skipping momenta are determined by degeneracies in the linearized scalar fluctuation equations. Both nevertheless encode the same near-horizon causal-area geometry, with the scalar spectrum providing additional information about the radial metric component.

Under suitable analyticity assumptions, the reconstruction can also extend beyond the small-velocity expansion. If the infinite pole-skipping tower is known and the reconstructed near-horizon metric series converges throughout the radial interval relevant to a given velocity, the blackening function \(f(r)\) can be determined in that region. The worldsheet-horizon position then follows from
\begin{equation}
v^2
=
\frac{f(r_s)}{r_s^2},
\label{eq:PSexactrs}
\end{equation}
and the corresponding finite-velocity force is
\begin{equation}
F(v)
=
-\sigma v\,r_s(v)^2 .
\label{eq:PSexactF}
\end{equation}
Once the normalized drag curve has been reconstructed, the exterior geometric membrane follows parametrically from \eqref{eq:qfromH} and \eqref{eq:MfromH}.

The extension from local to finite-velocity reconstruction is conditional on the analytic properties of the background. In particular, the near-horizon series must converge, or admit a controlled analytic continuation, up to the worldsheet horizon being probed. The reconstruction is consequently limited by the singularities and critical points of the analytically continued radial geometry, as in the general pole-skipping reconstruction program \cite{LuRanWu2026}. Without such assumptions, the pole-skipping hierarchy determines the local near-horizon data but need not uniquely fix the complete exterior geometry.

The correspondence also provides a spectral formulation of the null-energy constraints derived earlier. Since the NEC restricts the butterfly velocity and the local membrane geometry, the same restrictions must be reflected in the corresponding pole-skipping data. For a minimally coupled massless scalar, the butterfly-velocity bound \eqref{eq:vBbound} becomes
\begin{equation}
-\frac{\epsilon}{\kappa^2}
\geq
\frac{2p}{p+1}.
\label{eq:PSvBNEC}
\end{equation}
Thus, the first skipped momentum is constrained directly by the NEC through its relation to the butterfly scale.

At the next level, the spectral data determine the horizon value of the causal load introduced in \eqref{eq:Nload}. Using the first two pole-skipping levels, we obtain
\begin{equation}
\cN(0)
=
\frac{
4(3p+1)\epsilon^2
-4p\epsilon S_2
+pP_2
}{
8p\epsilon^2}.
\label{eq:N0PS}
\end{equation}
The requirement \(\cN(0)\geq0\) therefore implies
\begin{equation}
4(3p+1)\epsilon^2
-4p\epsilon S_2
+pP_2
\geq0 .
\label{eq:PSNECsecond}
\end{equation}
Similarly, regular convexity of the squared membrane tension requires
\begin{equation}
P_2-4\epsilon S_2+16\epsilon^2
>
0 .
\label{eq:PSconvexity}
\end{equation}

These inequalities show that the pole-skipping spectrum is constrained not only by the near-horizon fluctuation equations but also by the gravitational energy conditions governing the background. The NEC can therefore be translated into algebraic restrictions on discrete complex-momentum data of a boundary two-point function. This provides a spectral counterpart to the bounds on the butterfly velocity, the membrane tension, and the nonlinear drag coefficients derived in the preceding sections.

As an analytic consistency check, consider the planar AdS--Schwarzschild black brane in area-radius gauge,
\begin{equation}
f(r)
=
r^2-r^{1-p},
\qquad
\beta(r)=1,
\qquad
\kappa=\frac{p+1}{2}.
\label{eq:SchPSmetric}
\end{equation}
For a minimally coupled massless scalar, the first skipped momentum and the symmetric combinations of the two second-level roots are
\begin{equation}
\epsilon
=
-\frac{p(p+1)}{2},
\qquad
S_2
=
-2p(p+1),
\qquad
P_2
=
p(p-1)(p+1)^2.
\label{eq:SchPSdata}
\end{equation}
Equivalently, the two second-level momenta are
\begin{equation}
\mu_{2,\pm}
=
-(p+1)\left(p\pm\sqrt p\right).
\label{eq:SchPSroots}
\end{equation}
Substituting these expressions into the reconstruction formulas, together with the corresponding higher-order metric coefficient for \(c_3\), gives
\begin{equation}
c_1
=
\frac{2}{p+1},
\qquad
c_2
=
\frac{p+3}{(p+1)^2},
\qquad
c_3
=
\frac{2(p+2)(p+3)}{3(p+1)^3},
\label{eq:SchPScoeffcheck}
\end{equation}
in agreement with the exact binomial expansion \eqref{eq:Schcoefficients}.

The AdS--Schwarzschild example therefore verifies the consistency of the spectral reconstruction with the exact membrane--drag transformation and the nonlinear velocity expansion. It also provides a reference background for which the relevant NEC inequalities are saturated and the near-horizon geometry is known analytically.

An important qualification concerns the choice of scalar probe. The reconstruction developed above assumes a minimally coupled scalar with a specified mass. More general scalar actions may introduce additional radial functions into the fluctuation equation, modifying the relation between the skipped momenta and the background geometry.

To illustrate this, consider a scalar field with a radial kinetic coupling and an effective potential,
\begin{equation}
S_\Phi
=
-\frac{1}{2}
\int\dd^{p+2}x\sqrt{-g}\,
Z_\Phi(r)
\left[
(\partial\Phi)^2+U(r)\Phi^2
\right].
\label{eq:nonminimalscalar}
\end{equation}
The first skipped momentum then satisfies
\begin{equation}
\mu_{1,1}+U_h
=
-\frac{\kappa^2}{v_B^2}
\left[
1+\frac{1}{p}
\left(\log Z_\Phi\right)'_h
\right],
\label{eq:nonminimalfirstPS}
\end{equation}
in horizon units. The additional term involving the radial derivative of \(Z_\Phi\) shows that the pole-skipping momentum now contains operator-specific horizon information beyond the background metric. Consequently, the purely geometric reconstruction formulas cannot be applied without accounting for the coupling of the scalar probe.

Further complications arise when the scalar perturbation is not an independent probe but belongs to the active matter sector supporting the background. In such cases, scalar fluctuations can mix with longitudinal metric perturbations, and the near-horizon recursion must be formulated as a coupled block system rather than the single-field matrix \eqref{eq:recursionmatrix}. Although the organization of special frequencies follows from the near-horizon structure, the corresponding skipped momenta depend on the complete coupled fluctuation action \cite{WangWang2022}.

Finally, a degeneracy of the near-horizon recursion does not always guarantee a conventional pole-zero crossing in the boundary retarded Green function after ultraviolet matching. Anomalous and higher-type pole-skipping points require the more refined classification discussed in \cite{AhnEtAl2021}. For the reconstruction considered here, the essential input is therefore the set of roots of the near-horizon degeneracy determinant, together with a specified and appropriately calibrated scalar probe.

These qualifications clarify the scope of the spectral correspondence. For a minimally coupled scalar, the pole-skipping tower provides a systematic reconstruction of the near-horizon metric, from which the nonlinear drag coefficients and membrane endpoint derivatives follow. More general scalar couplings introduce additional operator-dependent information, analogous to the probe-frame dressing encountered in the trailing-string analysis. Once the probe action is specified, however, the same reconstruction framework continues to distinguish the geometric data shared by different observables from the additional information associated with their particular bulk realizations.

\section{Information recovery and the interior membrane geometry}
\label{sec:informationvelocity}

An interesting quantity characterizing the propagation and recovery
of quantum information in chaotic many-body systems is the information
velocity $v_I$ \cite{CouchEtAl2020,EcclesEtAl2021,
JiangMezeiVirrueta2025}. In quantum information theory, this velocity
characterizes the asymptotic growth of the smallest spatial region
from which quantum information initially encoded in a localized
degree of freedom can be recovered. Unlike the butterfly velocity,
which characterizes operator spreading, the information velocity
also depends on the entanglement structure of the initial state.
This dependence is conventionally described by a parameter $f$,
defined as the fraction of the final thermal entropy density
already present as initial volume-law entanglement.  
In particular, one may consider a strip-shaped information-recovery
protocol, in which the information velocity becomes a function of
the initial entanglement fraction $f$ and interpolates between the
entanglement and butterfly velocities,
$v_I(0)=v_E$ and $v_I(1)=v_B$.

Within the membrane description, the parameter $f$ has a particularly
simple mathematical interpretation. While the membrane tension
$\cE(u)$ is determined by the geometry of the final thermal state,
$f$ enters through the initial-state contribution to the membrane
variational problem. It therefore acts as a parameter specifying
the boundary conditions of the membrane, with the preferred slope
$u_f$ determined by $\cE'(u_f)=f$. This formulation makes it
possible to study the dependence of information propagation on
initial entanglement using the same geometric membrane tension
that appears in our correspondence with the stationary drag force.

Our purpose in this section is to explore the consequences of
this common geometric structure for information velocity and
the reconstruction of the black-hole interior. We first treat
$f$ as a formal parameter labeling a family of information-recovery
problems governed by a fixed thermal membrane, and derive its
relation to the nonlinear drag coefficients and the underlying
bulk geometry. At the same time, the physical interpretation
of this parameter in the specific heavy-quark setup considered
here deserves additional attention: unlike the quark velocity,
$f$ is not an independent parameter of the stationary
trailing-string solution. We therefore subsequently discuss
possible physical realizations of the initial entanglement
fraction within the same strongly coupled medium, emphasizing
a mechanism in which energy dissipated by heavy probes prepares
a state with a nontrivial entanglement deficit. These
considerations provide a natural physical setting for the
information-theoretic relations derived below, although we
do not attempt a complete dynamical treatment of the preparation
process or claim that the proposed interpretations exhaust all
possible realizations. Establishing their full holographic
description, including the relevant time-dependent backreaction
and entanglement dynamics, remains an interesting direction
for further investigation.

\subsection{Information velocity and the nonlinear drag expansion}
\label{subsec:informationdrag}

As discussed above, we consider the strip-shaped information-recovery
protocol introduced in \cite{CouchEtAl2020,EcclesEtAl2021,
JiangMezeiVirrueta2025}. Let us briefly recall its physical
formulation and the corresponding membrane description. The protocol
begins with quantum information initially encoded in a localized
degree of freedom entangled with an external reference system.
Under the subsequent many-body time evolution, this information
spreads through the system, and one asks how large a spatial region
must be accessible in order to recover it. The information velocity
characterizes the asymptotic rate at which the size of this
recovery region grows. An important feature of the construction
is its dependence on the entanglement already present in the
initial state, characterized by the fraction $f$ of the final
thermal entropy density.

For the strip realization, one considers a boundary region $A$
of width $2R$ and transverse volume $V_\perp$, in the large-region
and late-time scaling limit. The initial state is assumed to
possess a volume-law entanglement entropy density
$f s_{\mathrm{th}}$, where $s_{\mathrm{th}}$ is the entropy
density of the final thermal state. In the membrane description,
the dynamics is governed by the tension $\cE(u)$ associated with
this final state, while the initial entanglement enters through
an additional contribution at the lower endpoint of the membrane.
For a symmetric strip, the two membrane branches contribute
equally. A branch with slope $u$ extends over a time interval
$t$ and reaches the initial time slice at a distance $R-ut$
from the strip center. Its contribution consists of the
membrane cost $t\cE(u)$ and the initial-state entropy cost
$f(R-ut)$. Minimizing their sum over the allowed membrane
slopes gives the entropy on the unsaturated branch,

\begin{equation}
\frac{S_A(t)}{2s_{\mathrm{th}}V_\perp}
=
\min_{0\leq u\leq v_B}
\left[t\cE(u)+f(R-ut)\right] .
\label{eq:informationstripentropy}
\end{equation}
Here $u$ is the local slope of the physical membrane, which lies on the interior branch $0\leq u\leq v_B$. For a smooth strictly convex tension, the minimizing slope $u_f$ is fixed by
\begin{equation}
\cE'(u_f)=f .
\label{eq:informationstationarity}
\end{equation}
Saturation occurs when $S_A(t_{\mathrm{sat}})=2s_{\mathrm{th}}V_\perp R$. Identifying the information velocity with the large-strip saturation speed, $v_I(f)=R/t_{\mathrm{sat}}$, therefore gives
\begin{equation}
v_I(f)
=
\frac{\cE(u_f)-fu_f}{1-f},
\qquad
\cE'(u_f)=f,
\qquad 0\leq f<1 .
\label{eq:informationvelocitymembrane}
\end{equation}
The endpoint $f=1$ is understood as a limit. For the regular holographic strip branch considered here, the strict convexity of $\cE(u)$ and the endpoint conditions \eqref{eq:endpointconditions} give $u_f\to v_B$ as $f\to1^-$, while $u_f\to0$ as $f\to0^+$. Equation \eqref{eq:informationvelocitymembrane} consequently yields $v_I(0)=\cE(0)=v_E$ and $v_I(1)=v_B$. The strip geometry and the existence of the smooth minimizing branch are important: other region shapes, membrane transitions, or additional microscopic constraints can modify the corresponding information-recovery prescription.

The connection with stationary drag becomes particularly transparent for nearly thermal initial entanglement. Set
\begin{equation}
\delta=1-f,
\qquad
b=v_B,
\qquad
e_j=\cE^{(j)}(b) .
\label{eq:informationendpointdefinitions}
\end{equation}
Using $\cE(b)=b$ and $\cE'(b)=1$, the stationarity condition \eqref{eq:informationstationarity} gives
\begin{equation}
u_f
=
b-\frac{\delta}{e_2}
-\frac{e_3}{2e_2^3}\delta^2
+O(\delta^3) .
\label{eq:informationslopeexpansion}
\end{equation}
Expanding the numerator of \eqref{eq:informationvelocitymembrane} around $u=b$, and using this expression to eliminate $u_f$, we find
\begin{equation}
v_I(f)
=
b
-\frac{1-f}{2\cE''(b)}
-\frac{\cE'''(b)}{6\cE''(b)^3}(1-f)^2
+O\!\left((1-f)^3\right) .
\label{eq:informationendpointexpansion}
\end{equation}
The coefficient of the first correction is the inverse curvature of the membrane tension at the butterfly point. This is precisely the quantity determined by the $v^5$ term in the common-frame drag force. Substituting the relation \eqref{eq:c2E}, we obtain
\begin{equation}
 v_I(f)
 =
 v_B
 -\frac{p v_B^5}{2}
 \left[\frac{p+1}{2}c_1^2-c_2\right](1-f)
 +O\!\left((1-f)^2\right)
\label{eq:informationvelocitydrag}
\end{equation}
or, equivalently,
\begin{equation}
\left.\frac{\dd v_I}{\dd f}\right|_{f=1^-}
=
\frac{p v_B^5}{2}
\left[\frac{p+1}{2}c_1^2-c_2\right]
=
\frac{1}{2\cE''(v_B)}>0 .
\label{eq:informationendpointderivative}
\end{equation}
Thus the cubic drag coefficient fixes the limiting information velocity through $v_B^2=1/(pc_1)$, whereas the quintic coefficient fixes its first variation with the initial entanglement fraction. The next correction in \eqref{eq:informationendpointexpansion} depends on $\cE'''(v_B)$ and is determined by $c_3$ through \eqref{eq:c3E}. More generally, the triangular correspondence between drag coefficients and derivatives of the membrane tension provides a local expansion of the strip information velocity around $f=1$, order by order.

For completeness, this hierarchy can also be written explicitly through second order in $1-f$ using only the drag coefficients. Introducing
\begin{equation}
d_\star
=
\frac{p+1}{2p}-\frac{c_2}{pc_1^2}
=
\frac{1}{v_B\cE''(v_B)},
\label{eq:informationdstar}
\end{equation}
we find
\begin{equation}
\begin{aligned}
\frac{v_I(f)}{v_B}
={}&1-\frac{d_\star}{2}(1-f)
\\
&-\frac{(1-f)^2}{12}
\bigg[
\frac{6c_3}{p^2c_1^3}
-\left(1+\frac1p\right)\left(2+\frac1p\right)
+6\left(2+\frac1p\right)d_\star
-12d_\star^2
\bigg]
+O\!\left((1-f)^3\right) .
\end{aligned}
\label{eq:informationdragthroughc3}
\end{equation}
These identities concern the physical interior membrane in the vicinity of its horizon endpoint. They do not imply that a finite set of exterior drag coefficients determines $v_I(f)$ throughout $0\leq f\leq1$, since the latter generally requires the membrane farther inside the black hole.

\subsection{Reconstructing the interior causal-area geometry from information recovery}
\label{subsec:informationinterior}

The information-velocity relation also admits an inverse construction. Define
\begin{equation}
\Gamma(f)=(1-f)v_I(f)
=\cE(u_f)-fu_f .
\label{eq:informationGammadef}
\end{equation}
This is the Legendre transform of the physical membrane tension, with the sign convention adapted to the minimization in \eqref{eq:informationstripentropy}. Differentiating with respect to $f$ and using $\cE'(u_f)=f$, we obtain the envelope identity
\begin{equation}
\Gamma'(f)=-u_f .
\label{eq:informationGammaenvelope}
\end{equation}
Therefore, knowledge of the full function $v_I(f)$ reconstructs both the membrane slope and its tension parametrically:
\begin{equation}
\begin{aligned}
u_f
&=v_I(f)-(1-f)v_I'(f),\\
\cE(u_f)
&=v_I(f)-f(1-f)v_I'(f) .
\end{aligned}
\label{eq:informationinverseE}
\end{equation}
A necessary regularity condition for this construction is
\begin{equation}
\frac{\dd u_f}{\dd f}
=
2v_I'(f)-(1-f)v_I''(f)
=
\frac{1}{\cE''(u_f)}>0 ,
\label{eq:informationconvexity}
\end{equation}
which guarantees that the physical membrane slope increases monotonically with $f$.

We may now combine this Legendre inversion with the geometric reconstruction derived in \cref{sec:exacttransform}. Writing $q=u_f^2$ and $M(q)=\cE(u_f)^2$, the identity $M'(q)=Y$ implies
\begin{equation}
Y(f)
=
\frac{\cE(u_f)\cE'(u_f)}{u_f}
=
\frac{f\cE(u_f)}{u_f} .
\label{eq:informationinteriorY}
\end{equation}
Similarly, the causal function is
\begin{equation}
\begin{aligned}
a(f)
&=u_f^2-\frac{\cE(u_f)^2}{Y(f)}
=u_f^2-\frac{u_f\cE(u_f)}{f}
\\
&=-\frac{u_f(1-f)v_I(f)}{f} .
\end{aligned}
\label{eq:informationinteriora}
\end{equation}
Eliminating the auxiliary quantities $u_f$ and $\cE(u_f)$ with \eqref{eq:informationinverseE}, we arrive at the direct parametric reconstruction
\begin{equation}
\begin{aligned}
Y(f)
&=
\frac{f\big[v_I(f)-f(1-f)v_I'(f)\big]}
{v_I(f)-(1-f)v_I'(f)},
\\[2pt]
a(f)
&=-\frac{(1-f)v_I(f)}{f}
\big[v_I(f)-(1-f)v_I'(f)\big],
\qquad 0<f<1 .
\end{aligned}
\label{eq:informationinteriorreconstruction}
\end{equation}
For a regular strictly convex physical branch, these expressions determine the causal-area relation $a(Y)$ from the horizon, where $(Y,a)=(1,0)$, down to the entanglement bottleneck at $q=0$. This is complementary to \eqref{eq:reducedtomography}: physical stationary drag reconstructs $a(Y)$ in the exterior, whereas the full strip information-velocity curve reconstructs it on the interior interval sampled by the entanglement membrane. Both descriptions meet at the butterfly point. Neither determines the independent radial lapse, nor does the information-velocity construction extend by itself beyond the entanglement bottleneck into deeper regions of the interior.

The two endpoints provide useful consistency conditions. At $f\to1^-$, the membrane endpoint relations give
\begin{equation}
v_I(1)=v_B,
\qquad
Y(1)=1,
\qquad
a(1)=0 ,
\label{eq:informationhorizonmatching}
\end{equation}
where the first derivative $v_I'(1^-)$ is fixed by the nonlinear drag coefficient $c_2$ through \eqref{eq:informationendpointderivative}. At $f\to0^+$, isotropy implies $u_f\to0$ and $\cE'(0)=0$, so the first equation in \eqref{eq:informationinverseE} requires
\begin{equation}
v_I(0)=v_I'(0)=v_E .
\label{eq:informationbottleneckcondition}
\end{equation}
If $\cE''(0)$ is finite and positive, then
\begin{equation}
2v_E-v_I''(0)
=\frac{1}{\cE''(0)},
\label{eq:informationbottleneckcurvature}
\end{equation}
and the limiting values of \eqref{eq:informationinteriorreconstruction} determine the bottleneck geometry,
\begin{equation}
Y_E=
\frac{v_E}{2v_E-v_I''(0)},
\qquad
a_E=-v_E\big[2v_E-v_I''(0)\big] .
\label{eq:informationbottleneckgeometry}
\end{equation}
Thus, the value and curvature of the information-velocity curve near the initially unentangled endpoint determine, respectively, the entanglement velocity and the causal-area location of the interior bottleneck. The reconstruction is based on the family of initial states entering the same final-state strip membrane theory; it should not be applied to information-velocity curves obtained from unrelated final geometries.

\subsection{An illustration in AdS$_5$--Schwarzschild}
\label{subsec:informationexample}

As a simple analytic check, consider the planar AdS$_5$--Schwarzschild background, for which $p=3$ and $v_B^2=2/3$. The squared membrane tension in \eqref{eq:powerlawM} gives
\begin{equation}
\cE(u)
=\frac{\sqrt{2}}{3^{3/4}}(1-u^2)^{-1/4},
\qquad 0\leq u\leq\sqrt{\frac23} .
\label{eq:informationSchmembrane}
\end{equation}
Its derivative, $f=\cE(u)u/[2(1-u^2)]$, together with \eqref{eq:informationvelocitymembrane}, defines the full strip information velocity parametrically. Using the coefficients $c_1=1/2$, $c_2=3/8$, and $c_3=5/16$ from \eqref{eq:AdS5coefficients}, the endpoint expansion becomes
\begin{equation}
 v_I(f)
 =\sqrt{\frac23}\left[
 1-\frac{1-f}{12}
 -\frac{5}{108}(1-f)^2
 +O\!\left((1-f)^3\right)
 \right] .
\label{eq:informationSchendpoint}
\end{equation}
At the opposite endpoint, $v_E=\sqrt{2}/3^{3/4}$ and $\cE''(0)=v_E/2$, so \eqref{eq:informationbottleneckgeometry} gives
\begin{equation}
Y_E=3^{-3/2},
\qquad a_E=-2 .
\label{eq:informationSchbottleneck}
\end{equation}
These are precisely the values obtained directly from $a(z)=1-z^4$ and $Y=z^{-6}$ at the Schwarzschild entanglement bottleneck $z^4=3$. This example verifies both the near-butterfly drag expansion and the interior reconstruction from the information-velocity function.

The two results developed in this section clarify the complementary roles of the observables. The nonlinear drag expansion fixes the local behavior of information recovery near maximal initial entanglement, while the full information-velocity curve supplies a real-domain reconstruction of the physical interior membrane geometry. The comparison is therefore strongest when the exterior response and the family of information-recovery protocols are evaluated in the same thermal background. In the absence of additional analyticity or dynamical assumptions, the exterior force alone does not determine the complete interior information-velocity function.

\subsection{What entanglement fraction $f$ means in the drag force setup}
\label{subsec:informationfraction}

The connection between nonlinear drag and information velocity established
above raises a natural question concerning the physical meaning of the
initial entanglement fraction $f$ in the present setting. The stationary
trailing-string solution describes the dissipative motion of a heavy quark
through an equilibrium thermal medium, whereas the information velocity
depends on the amount of entanglement already present in the state before
the information-recovery process begins. At first sight, these quantities
appear to characterize different physical situations, since the parameter
$f$ does not enter the stationary string action or the worldsheet-horizon
condition. Nevertheless, there is a natural way to relate them within
the same physical system. The energy transferred from a moving heavy
quark to the plasma provides a mechanism for preparing a state with a
nontrivial initial entanglement fraction, while the final thermal
geometry determines the subsequent propagation and recovery of
quantum information.

The underlying mechanism relies on the distinction between the local
thermalization of energy and the equilibration of entanglement over
large spatial regions. A strongly coupled plasma can redistribute
deposited energy among its local degrees of freedom on timescales
much shorter than those required for entanglement to equilibrate
across a sufficiently large region. Consequently, the local
thermodynamic properties may already approach those of a new
equilibrium state, while the leading volume-law entanglement
entropy still retains information about the earlier state.
This separation naturally produces an initial entanglement
fraction smaller than unity when the final equilibrium
entropy density exceeds the entropy density associated
with the initial state.

The heavy-quark drag mechanism considered in this work provides
a direct source of such energy deposition. A quark maintained at
constant velocity $v$ experiences the drag force
$F(v)=-\sigma vH(v^2)$. The external work required to sustain
its motion is transferred to the medium, with the corresponding
power given by
\begin{equation}
P(v)=-vF(v)=\sigma v^2H(v^2).
\label{eq:informationdragwork}
\end{equation}
In the holographic description, this energy transfer is
represented by the conserved radial energy flux carried
by the trailing string toward the black-brane horizon
\cite{HerzogEtAl2006,Gubser2006}. Thus, the same
worldsheet configuration that determines the stationary
momentum loss also provides the mechanism by which
mechanical work is deposited into the thermal system.

To make the preparation approximately homogeneous, one may
consider a spatially distributed ensemble of heavy probes,
with oppositely directed velocities so that the net
momentum injection vanishes while the deposited energies
add. For a total probe density $n_Q$, the power transferred
to the plasma per unit volume is
\begin{equation}
\mathcal{P}(t)
=
n_Q\sigma(T(t))v(t)^2H_{T(t)}(v(t)^2).
\label{eq:informationdragpower}
\end{equation}
The temperature dependence allows for the gradual evolution
of the thermodynamic state during preparation. We assume
that this evolution is sufficiently slow compared with
the local relaxation of the string and plasma, so that
the stationary drag formula can be applied quasistatically.

If the deposited energy remains in the plasma, its
energy density evolves according to
\begin{equation}
\frac{\dd\varepsilon}{\dd t}
=
\mathcal{P}(t).
\label{eq:informationenergyevolution}
\end{equation}
At fixed conserved charge densities, local thermodynamic
equilibrium implies
\begin{equation}
T\frac{\dd s_{\mathrm{eq}}}{\dd t}
=
\mathcal{P}(t),
\label{eq:informationentropyevolution}
\end{equation}
where $s_{\mathrm{eq}}$ denotes the equilibrium entropy
density associated with the instantaneous local energy
density. The energy supplied by the heavy probes therefore
increases the equilibrium entropy density of the plasma.

Suppose that the initial state has equilibrium entropy
density $s_i$, and that the driving is terminated after
the accumulated deposited work density
\begin{equation}
\mathcal{W}
=
\int_{\mathrm{prep}}\dd t\,\mathcal{P}(t)
\label{eq:informationaccumulatedwork}
\end{equation}
has increased the energy density from $\varepsilon_i$
to $\varepsilon_f=\varepsilon_i+\mathcal{W}$.
The corresponding final equilibrium entropy density
is $s_f=s_{\mathrm{eq}}(\varepsilon_f)$.
If the large-region entanglement has not yet equilibrated
to this new value, its initial fraction relative to the
final thermal state is naturally
\begin{equation}
f\simeq\frac{s_i}{s_f}.
\label{eq:informationdragpreparedf}
\end{equation}
In this interpretation, the parameter $f$ is not imposed
independently of the heavy-quark dynamics. It is generated
by the dissipative preparation of the plasma and is
determined by the accumulated work transferred through
the trailing strings.

The relation can be expressed more explicitly using the
equation of state. For a neutral conformal plasma in $p$
spatial dimensions, one has
$\varepsilon\propto T^{p+1}$ and
$s_{\mathrm{eq}}\propto T^p$. Consequently,
\begin{equation}
f\simeq
\left(
\frac{\varepsilon_i}{\varepsilon_f}
\right)^{p/(p+1)}
=
\left(
\frac{\varepsilon_i}
{\varepsilon_i+\mathcal{W}}
\right)^{p/(p+1)}.
\label{eq:informationdragconformalf}
\end{equation}
This expression makes the physical origin of the fraction
particularly transparent. In the absence of energy deposition,
$\mathcal{W}=0$ and $f=1$, while positive deposited work produces
$f<1$. The amount of initial entanglement relative to the final
thermal entropy is therefore controlled by the energy dissipated
during the heavy-probe motion.

Equivalently, during the preparation one may define the
instantaneous ratio
$f(t)=s_i/s_{\mathrm{eq}}(t)$. Combining this definition with
\eqref{eq:informationentropyevolution} gives
\begin{equation}
\frac{\dd f}{\dd t}
=
-f\,
\frac{
n_Q\sigma(T)v^2H_T(v^2)
}{
T s_{\mathrm{eq}}(T)
}.
\label{eq:informationdragfevolution}
\end{equation}
The positivity of the dissipated power implies that this
ratio decreases during heating. In this sense, the drag
force directly controls the preparation of the entanglement
deficit relative to the instantaneous equilibrium entropy.
The identification of this thermodynamic ratio with the
initial entanglement fraction entering the membrane theory,
however, applies when the actual large-region entanglement
has remained approximately at its initial value throughout
the preparation.

This condition is realized when the relevant timescales are
well separated. Denoting by $\tau_{\mathrm{drag}}$ and
$\tau_{\mathrm{loc}}$ the relaxation times of the probe
and the local plasma, respectively, and by
$\tau_{\mathrm{prep}}$ the duration of the driving, a useful
regime for a sufficiently large strip of half-width $R$ is
\begin{equation}
\tau_{\mathrm{drag}},\tau_{\mathrm{loc}}
\ll \tau_{\mathrm{prep}}
\ll \frac{R}{v_E},
\label{eq:informationpreparationtimescales}
\end{equation}
where $v_E$ denotes the characteristic entanglement-growth
velocity of the final thermal medium. The first inequality
allows the drag force to be treated quasistatically, whereas
the second ensures that the large-region entanglement
need not reach its new equilibrium value during preparation.
Under these assumptions, and provided the driving does not
introduce an additional uncontrolled extensive entanglement
contribution, the local thermal properties can be
characterized by the final state while the leading
volume-law entanglement density remains approximately $s_i$.

The interpretation of $f$ in the membrane variational
problem is then immediate. The final thermal geometry
determines the membrane tension $\cE(u)$ and the
normalization $s_f$, while the entanglement retained
from the earlier state supplies the initial boundary
contribution proportional to $s_i$. Expressing this
contribution in units of the final thermal entropy
density produces precisely the factor $f=s_i/s_f$
appearing in \eqref{eq:informationstripentropy}.
The information velocity subsequently characterizes
the propagation and recovery of quantum information
in the final medium, starting from the entanglement
structure prepared by heavy-probe dissipation.

This construction also provides a direct quantitative
interpretation of the relation derived in
\eqref{eq:informationvelocitydrag}. For weak heating,
$\mathcal{W}/\varepsilon_f\ll1$, the conformal result
\eqref{eq:informationdragconformalf} gives
\begin{equation}
1-f
=
\frac{p}{p+1}
\frac{\mathcal{W}}{\varepsilon_f}
+
O\left(
\frac{\mathcal{W}^2}{\varepsilon_f^2}
\right).
\label{eq:informationweakheatingf}
\end{equation}
Substituting into the near-butterfly expansion of
the information velocity, we obtain
\begin{equation}
\begin{aligned}
v_B-v_I(f)
={}&
\frac{p^2v_B^5}{2(p+1)}
\left[
\frac{p+1}{2}c_1^2-c_2
\right]
\frac{\mathcal{W}}{\varepsilon_f}
\\
&+
O\left(
\frac{\mathcal{W}^2}{\varepsilon_f^2}
\right).
\end{aligned}
\label{eq:informationdragworkvelocity}
\end{equation}
Here $v_B$, $c_1$, and $c_2$ are evaluated in the final
equilibrium background. The relation shows that the
work dissipated by the heavy probes determines the
initial entanglement deficit, whereas the nonlinear
drag coefficients determine the leading response of
information velocity to that deficit. The same physical
mechanism therefore enters both the preparation of the
initial state and the subsequent characterization of
information propagation.

An important distinction concerns the backgrounds used
in this comparison. The derivative of $v_I(f)$ with
respect to $f$ considered above is defined at fixed
final membrane tension. A family of preparations with
different values of $f$ can be compared within the same
final thermal geometry by appropriately varying the
initial thermodynamic state and the deposited work.
If instead the initial state is kept fixed while the
amount of work is varied, the final temperature and
generally the membrane tension change as well. The
information velocity then depends on both the prepared
fraction and the changing final geometry, and the
fixed-background derivative in
\eqref{eq:informationendpointderivative} should not
be interpreted as the total response to the drive.

The proposed interpretation thus naturally separates
the preparation stage from the subsequent information
dynamics. During preparation, the heavy probes transfer
energy to the plasma, and the stationary drag expression
supplies the local dissipative response in a quasistatic
approximation. After the driving has ceased and the
local medium has equilibrated, the resulting final
black-brane geometry determines the membrane tension
entering the information-recovery calculation. A finite
change of the background energy density necessarily
involves backreaction, so the complete preparation
process lies beyond the strictly stationary probe
solution. Nevertheless, the energy balance and the
separation of local and large-scale equilibration
provide a concrete physical mechanism for generating
the fraction $f$ within the same heavy-quark--plasma
system.

In this way, the relation between nonlinear drag and
information velocity acquires a unified dynamical
interpretation. The trailing-string energy flux
prepares a plasma whose available equilibrium entropy
exceeds its already established large-scale entanglement,
and the geometric membrane of the resulting thermal
state governs the subsequent evolution of that
entanglement deficit. The initial fraction $f$
therefore characterizes a state produced through
heavy-probe dissipation, rather than an unrelated
parameter added to the stationary drag calculation.

There are also other, less direct ways of assigning a meaningful
information-theoretic parameter to the same thermal medium. Unlike
the drag-induced preparation discussed above, these possibilities
do not establish an explicit causal connection between heavy-probe
energy loss and the initial entanglement fraction. Rather, they
provide complementary perspectives on how information-theoretic
data may be associated with a fixed thermal geometry. We briefly
outline these interpretations without attempting to establish
their complete dynamical realization or claiming that they are
equivalent to the original information-recovery protocol. Their
purpose is to illustrate the broader physical scope of the
membrane description and to indicate possible extensions of
the present framework.

\paragraph{Moving entanglement cut.}
A stationary thermal geometry can support nontrivial
information dynamics even in the absence of a change in
temperature or an initial-state quench. For example, one
may consider the entanglement of the thermal time-evolution
operator and choose planar entangling cuts displaced by a
distance $\ell$. In the membrane regime, their leading
operator entanglement is determined by the same tension,
$S_{\mathrm{op}}(t,\ell)\simeq
s_{\mathrm{th}}V_\perp t\cE(\ell/t)$.
Its normalized derivative with respect to the displacement
defines a conjugate information variable,
\begin{equation}
f_{\mathrm{cut}}
=
\frac{1}{s_{\mathrm{th}}V_\perp}
\frac{\partial S_{\mathrm{op}}}{\partial\ell}
=
\cE'(\ell/t).
\label{eq:informationmovingcut}
\end{equation}
If the displacement of the cut is chosen to follow the quark
trajectory, $\ell=vt$, this gives
$f_{\mathrm{cut}}=\cE'(v)$ on the physical membrane branch
$0\leq v\leq v_B$. This construction associates the conjugate
membrane variable with a natural observation geometry tied
to the moving probe. Its interpretation is nevertheless
distinct from the initial entanglement fraction: it measures
the response of operator entanglement to the displacement
of a cut, rather than the entropy already present in a
prepared initial state. The relation to information velocity
arises because both quantities involve the same Legendre
structure of $\cE(u)$.

\paragraph{Locally thermal initial states.}
A second possibility is to consider a family of states
whose local observables have already equilibrated to the
same thermal values, but whose entanglement over large
spatial regions differs. Such states may have approximately
the same local stress tensor and short-distance response
functions while carrying different volume-law entanglement
densities,
$S_A\simeq f s_{\mathrm{th}}\operatorname{Vol}(A)$.
Provided the probe-sector dynamics has also locally
equilibrated, the leading heavy-quark drag is then governed
by the same equilibrium black-brane geometry, whereas
the different values of $f$ specify different boundary
conditions for the entanglement membrane. This scenario
emphasizes that local thermal equilibrium does not by
itself imply equilibration of entanglement on arbitrarily
large scales. Unlike the drag-induced preparation described
above, however, the initial fraction is here independently
specified by the preparation of the many-body state, rather
than generated by the mechanical work of the heavy probes.

\paragraph{Recorded thermal preparations.}
A related interpretation follows from the fact that the
same thermal density matrix can admit different recorded
ensembles of pure states,
\begin{equation}
\rho_\beta
=
\sum_k p_k
|\psi_k\rangle\langle\psi_k|.
\label{eq:informationrecordedensemble}
\end{equation}
The individual states $|\psi_k\rangle$ may possess
different spatial entanglement properties, even though
the density matrix obtained after discarding the
preparation record is identical. Consequently, the
averaged drag force is the same for all such
decompositions, since the unconditioned response
depends only on $\rho_\beta$ and the prescribed
probe dynamics. If the preparation record is retained,
the average entropy of the conditioned states can
instead define a fraction $f$ through its leading
large-region volume-law contribution. This gives an
information-theoretic realization of a variable
initial entanglement fraction without changing the
unconditioned thermal plasma. Its use in the membrane
recovery prescription, however, requires a suitable
family of conditioned states admitting the same
coarse-grained membrane description.

These interpretations highlight the distinction between
the geometric properties of the medium and the
information-theoretic conditions under which it is
probed. The membrane tension is fixed by the thermal
geometry, while the variable $f$ characterizes either
the preparation of the many-body state or, in the
moving-cut construction, a conjugate entanglement
observable. Among these possibilities, drag-induced
preparation gives the most direct dynamical
realization in the present context. The energy flux
of the trailing string supplies the mechanism that
creates an initial entanglement deficit, and the
same final thermal geometry determines how that
deficit influences the propagation and recovery of
quantum information. The correspondence between drag
and information velocity can therefore be understood
as relating the dissipative preparation and subsequent
information dynamics of the same strongly coupled
medium, rather than as a comparison of unrelated
physical systems.

\section{Completing the exterior geometry from stochastic heavy-quark dynamics}
\label{sec:stochasticreconstruction}

The exact membrane--drag transformation reconstructs the causal-area relation $a(Y)$ but leaves the radial metric component undetermined, as emphasized in \eqref{eq:reducedmetric}. This limitation reflects the fact that the stationary momentum flux of the trailing string is fixed by regularity at its induced horizon and does not depend on the radial proper-distance element. The fluctuations of the same string provide complementary information. In particular, the absolute momentum-broadening coefficients depend on the Hawking temperature of the induced worldsheet horizon, which is sensitive to the radial metric function \cite{GiecoldIancuMueller2009,GiataganasSoltanpanahi2014}. We now show that, for a common-frame Nambu--Goto probe, the complete mean drag curve together with one absolute momentum-broadening function determines the entire exterior metric, up to an overall normalization. The second broadening function is then fixed and supplies an independent consistency check.

\subsection{Worldsheet temperature and radial proper distance}
\label{subsec:stochasticradial}

We first restrict to the same regular, isotropic, nonextremal exterior branch and constant effective string coupling $Z$ used in the geometric membrane--drag correspondence. On this branch, $a(r)$ is monotone, so it may be employed as a radial coordinate. To distinguish this radial coordinate from the velocity of any individual quark, we write
\begin{equation}
x=a(r),\qquad w=v^2,\qquad 0<x<1,
\label{eq:stochasticxw}
\end{equation}
and define the positive radial metric function $\mathcal B_x(x)$ through
\begin{equation}
\dd s_E^2
=-C(x)x\,\dd t^2+\mathcal B_x(x)\,\dd x^2
+C(x)\,\dd\bm{x}_p^2,
\qquad C(x)=C_hH(x).
\label{eq:stochasticmetricx}
\end{equation}
Here $H(x)$ is the normalized drag function reconstructed in \cref{sec:exacttransform}, and the spacetime horizon is located at $x=0$. The unknown function $\mathcal B_x(x)$ is precisely the remaining metric datum that cannot be obtained from stationary drag or the leading geometric membrane.

For a quark moving in the $x^1$ direction with squared velocity $w$, the stationary string embedding is $x^1(t,x)=vt+\xi(x)$. The determinant of its induced Einstein-frame metric becomes
\begin{equation}
-\det h^{(E)}
=C(x)(x-w)\mathcal B_x(x)+C(x)^2x\,\xi'(x)^2,
\label{eq:stochasticdet}
\end{equation}
where the prime in this subsection denotes differentiation with respect to $x$. The conserved momentum flux is fixed by regularity at the worldsheet horizon $x=w$ to be $\Pi_\xi=vZC(w)$. Eliminating $\xi'$ in favor of this flux and making a stationary redefinition of the worldsheet time coordinate to remove its off-diagonal metric component, we obtain
\begin{equation}
\dd s_{\mathrm{ws}}^2
=-C(x)(x-w)\,\dd\widehat t^{\,2}
+\mathcal B_x(x)
\frac{C(x)^2x}{C(x)^2x-wC(w)^2}\,\dd x^2.
\label{eq:stochasticwsdiag}
\end{equation}
The derivative of the denominator at $x=w$ introduces the dimensionless combination
\begin{equation}
\mathscr R(w)
=\left.\frac{\dd[xH(x)^2]/\dd x}{H(x)^2}\right|_{x=w}
=1+2w\frac{H'(w)}{H(w)}.
\label{eq:stochasticRdef}
\end{equation}
Consequently, the near-horizon induced metric has the Rindler form
\begin{equation}
\dd s_{\mathrm{ws}}^2
=-C(w)(x-w)\,\dd\widehat t^{\,2}
+\frac{\mathcal B_x(w)w}{\mathscr R(w)(x-w)}\,\dd x^2
+\cdots,
\label{eq:stochasticwsRindler}
\end{equation}
where the omitted terms are regular in $x-w$. Requiring regularity of its Euclidean continuation determines the worldsheet Hawking temperature measured with respect to the boundary-normalized time,
\begin{equation}
T_{\mathrm{ws}}(w)^2
=\frac{C_hH(w)\mathscr R(w)}
{16\pi^2w\mathcal B_x(w)}.
\label{eq:stochasticTwsmetric}
\end{equation}
Unlike the momentum flux, the worldsheet temperature therefore depends explicitly on the radial proper-distance function. Equation \eqref{eq:stochasticTwsmetric} is the geometric ingredient that allows stochastic dynamics to complete the exterior reconstruction.

\subsection{Reconstructing the full metric from drag and momentum broadening}
\label{subsec:stochasticinverse}

Let $\kappa_\perp(v)$ and $\kappa_\parallel(v)$ denote the zero-frequency transverse and longitudinal momentum-broadening coefficients of the heavy quark, with the standard normalization of the local Langevin force correlator. For brevity, when their argument is $w=v^2$ we write $\kappa_a(w)\equiv\kappa_a(v=\sqrt w)$. At leading classical string order, the horizon relations quoted in \eqref{eq:Langevincoefficients} take the form
\begin{equation}
\begin{aligned}
\kappa_\perp(w)
&=2T_{\mathrm{ws}}(w)\,\sigma H(w),\\
\kappa_\parallel(w)
&=2T_{\mathrm{ws}}(w)\,\sigma
\bigl[H(w)+2wH'(w)\bigr].
\end{aligned}
\label{eq:stochasticbroadening}
\end{equation}
Here $\sigma=T_fZC_h$ is the linear drag normalization introduced in \eqref{eq:sigmadef}. The temperature entering these formulas is the induced worldsheet temperature rather than, in general, the heat-bath temperature $T$.

The first equation in \eqref{eq:stochasticbroadening} determines the worldsheet temperature directly from the mean force and one absolute noise function,
\begin{equation}
T_{\mathrm{ws}}(w)
=\frac{\kappa_\perp(w)}{2\sigma H(w)}.
\label{eq:stochasticTwsinverse}
\end{equation}
Substituting this expression into \eqref{eq:stochasticTwsmetric} gives
\begin{equation}
\mathcal B_x(x)
=\frac{C_h\sigma^2H(x)^3\mathscr R(x)}
{4\pi^2x\kappa_\perp(x)^2}.
\label{eq:stochasticBxreconstructed}
\end{equation}
As a result, the full exterior metric can be written solely in terms of the normalized mean drag, its first derivative, and the absolute transverse broadening,
\begin{equation}
\begin{aligned}
\dd s_E^2
=C_hH(x)\bigg[
&-x\,\dd t^2+\dd\bm{x}_p^2\\
&+\frac{\sigma^2H(x)^2}{4\pi^2x\kappa_\perp(x)^2}
\left(1+2x\frac{H'(x)}{H(x)}\right)\dd x^2
\bigg].
\end{aligned}
\label{eq:stochasticfullmetric}
\end{equation}
This reconstruction holds pointwise throughout the radial interval sampled by the worldsheet horizons, $0<x<1$. It assumes the common-frame string action, a regular monotone exterior branch, and the validity of the low-frequency classical worldsheet fluctuation relations. It determines the metric up to the overall factor $C_h$, which may be calibrated from the entropy density or, if $T_fZ$ is known, from $\sigma=T_fZC_h$.

To express the same result in the transverse-area coordinate used in \eqref{eq:reducedmetric}, recall that $Y=H(x)^p$ and $x=a(Y)$. Under this change of radial coordinate,
\begin{equation}
\mathcal B_Y(Y)
=\frac{C_hY^{1/p}}{16\pi^2a(Y)T_{\mathrm{ws}}(a(Y))^2}
\left[1+\frac{2a(Y)}{pY a_Y(Y)}\right]
a_Y(Y)^2,
\qquad a_Y=\frac{\dd a}{\dd Y}.
\label{eq:stochasticBYreconstructed}
\end{equation}
Thus the function $\mathcal B_Y(Y)$ that was left undetermined by the exact membrane--drag transformation is fixed once the absolute stochastic response is supplied. No use of the bulk Einstein equations is required in obtaining \eqref{eq:stochasticfullmetric} or \eqref{eq:stochasticBYreconstructed}; those equations may instead be imposed afterward as consistency conditions on the reconstructed geometry.

The longitudinal broadening provides a further prediction rather than an additional unknown. Dividing the two equations in \eqref{eq:stochasticbroadening} yields
\begin{equation}
\frac{\kappa_\parallel(x)}{\kappa_\perp(x)}
=\mathscr R(x)
=1+2x\frac{H'(x)}{H(x)},
\label{eq:stochasticratioidentity}
\end{equation}
reproducing the ratio in \eqref{eq:Langevinratio}. Conversely, the ratio of the two broadening functions determines the normalized mean drag even if the force has not been supplied independently:
\begin{equation}
H(x)
=\exp\left[
\frac12\int_0^x\frac{\dd y}{y}
\left(\frac{\kappa_\parallel(y)}{\kappa_\perp(y)}-1\right)
\right],
\qquad H(0)=1.
\label{eq:stochasticHfromratio}
\end{equation}
At zero velocity $T_{\mathrm{ws}}(0)=T$, so that $\kappa_\perp(0)=2\sigma T$. Accordingly, the two absolute noise curves together with the bath temperature determine $\sigma$, $H(x)$, and the radial metric coefficient, up to the overall geometric normalization. The independently computed mean drag then tests the joint worldsheet description.

The null energy condition imposes a further restriction on this stochastic reconstruction. In the Einstein--NEC class studied in \cref{sec:NEC}, the causal stiffness obeys $\Delta(x)\le B_p(1-x)$, whereas \eqref{eq:stochasticRdef} and \eqref{eq:Deltaequiv} imply $\mathscr R(x)=1+2x/[p\Delta(x)]$. It follows that
\begin{equation}
\frac{\kappa_\parallel(x)}{\kappa_\perp(x)}
\ge 1+\frac{4x}{(p+1)(1-x)}.
\label{eq:stochasticNECratio}
\end{equation}
For $p=3$, this reduces to $\kappa_\parallel/\kappa_\perp\ge (1-v^2)^{-1}$. This inequality is specific to the common-frame Einstein--NEC setting; a radially dressed string does not in general obey the same bound on its physical broadening ratio.

\subsection{AdS$_5$--Schwarzschild check and probe-frame dependence}
\label{subsec:stochasticexample}

The reconstruction can be verified directly in the planar AdS$_5$--Schwarzschild background, for which $p=3$ and
\begin{equation}
H(x)=(1-x)^{-1/2},\qquad
\mathscr R(x)=(1-x)^{-1}.
\label{eq:stochasticSchH}
\end{equation}
The induced worldsheet temperature and transverse broadening are
\begin{equation}
T_{\mathrm{ws}}(x)=T(1-x)^{1/4},\qquad
\kappa_\perp(x)=2\sigma T(1-x)^{-1/4}.
\label{eq:stochasticSchnoise}
\end{equation}
Substitution into \eqref{eq:stochasticfullmetric} gives
\begin{equation}
\dd s_E^2
=C_h(1-x)^{-1/2}\left[
-x\,\dd t^2+\dd\bm{x}_3^2
+\frac{\dd x^2}{16\pi^2T^2x(1-x)^{3/2}}
\right].
\label{eq:stochasticSchmetric}
\end{equation}
Introducing the standard Schwarzschild radial coordinate by $z/z_h=(1-x)^{1/4}$ and using $T=1/(\pi z_h)$, $C_h=L^2/z_h^2$, we recover
\begin{equation}
\dd s_E^2
=\frac{L^2}{z^2}
\left[
-\left(1-\frac{z^4}{z_h^4}\right)\dd t^2
+\frac{\dd z^2}{1-z^4/z_h^4}
+\dd\bm{x}_3^2
\right].
\label{eq:stochasticSchstandard}
\end{equation}
The longitudinal broadening is simultaneously fixed to be $\kappa_\parallel(x)=2\sigma T(1-x)^{-5/4}$, and the NEC inequality \eqref{eq:stochasticNECratio} is saturated at every velocity, as expected for this reference geometry.

Finally, the restriction to constant $Z$ is essential if the reconstructed metric is to be interpreted as the Einstein-frame geometry. For the general dressed string action \eqref{eq:stringaction}, the Nambu--Goto functional may instead be regarded as the ordinary string area in the effective metric $g^{(s)}_{MN}=Z(r)g^{(E)}_{MN}$. The same stochastic inversion, with $H$ replaced by the physical dressed function $H_Z$ and $\mathscr R$ by $\mathscr R_Z=1+2xH_Z'/H_Z$, reconstructs this effective string-frame metric. Without independent knowledge of the dressing function $Z(r)$, it does not uniquely reconstruct the Einstein-frame metric or the geometric entanglement membrane. This distinction is the stochastic counterpart of the probe-frame limitation established in \cref{sec:dressing}.

The combined deterministic and stochastic observables therefore resolve the information deficit identified in the original membrane--drag correspondence: stationary momentum loss determines the causal-area geometry, while the absolute worldsheet fluctuations determine the missing radial proper distance. In this sense, a velocity scan of mean drag and momentum broadening supplies a complete exterior geometric reconstruction within the specified classical probe theory, whereas its extension to the black-hole interior or to independent microscopic matter couplings requires additional information.

\section{Final comments and conclusion}
\label{sec:further}

The geometric structure developed in the preceding sections also provides a useful perspective on several related aspects of holographic dynamics. In particular, the membrane--drag correspondence connects characteristic information velocities, nonlinear momentum loss, and certain properties of worldsheet fluctuations through the same reduced causal-area geometry. At the same time, it makes explicit which aspects of the bulk geometry are accessible through this correspondence and where additional information is required. We discuss these points below, together with two special situations that help clarify the scope of the construction.

We first note that the Legendre structure of the geometric membrane implies a simple relation between the entanglement and butterfly velocities. Recall that the entanglement velocity is determined by the membrane tension at the interior bottleneck, \(v_E^2=M(0)\), whereas the butterfly point satisfies \(M(v_B^2)=v_B^2\). Integrating the identity \(M'(q)=Y(q)\) between these two points gives
\begin{equation}
v_B^2-v_E^2
=
\int_0^{v_B^2}
Y(q)\,\dd q .
\label{eq:vBvEsumrule}
\end{equation}
Since the normalized transverse-area density is positive, this relation immediately implies the familiar inequality \(v_E\leq v_B\). More importantly, it provides a geometric interpretation of the difference between the two velocities: their squared separation is given by the area under the interior branch of the canonical curve \(Y(q)\). The entanglement and butterfly velocities are therefore related through the integrated geometry of the membrane between its bottleneck and the butterfly point.

A related observation concerns the connection between nonlinear drag and the near-horizon scale governing quantum chaos. Introducing
\begin{equation}
D_B
=
\frac{v_B^2}{\kappa},
\qquad
\kappa=2\pi T,
\label{eq:DBdef}
\end{equation}
the leading nonlinear drag coefficient in the common-frame construction can be expressed as
\begin{equation}
c_1
=
\frac{1}{p\kappa D_B}.
\label{eq:c1DB}
\end{equation}
The same relation can be written in terms of the complex momentum of the energy-density pole-skipping point. Using \(k_\chi=i\kappa/v_B\), one obtains
\begin{equation}
c_1
=
-\frac{k_\chi^2}{p\kappa^2}.
\label{eq:c1chaosPS}
\end{equation}
Thus, the first nonlinear correction to the stationary drag force is determined by the same near-horizon scale that controls the complex-momentum location of the chaos-related energy-density pole-skipping point. This connection is particularly interesting because the two quantities characterize different physical processes: the drag coefficient describes the steady-state response of a moving external quark, whereas the pole-skipping momentum is associated with the analytic structure of a boundary retarded correlator.

The relation is nevertheless limited to the geometric information fixed by the butterfly velocity. Higher nonlinear drag coefficients probe the curvature and successive derivatives of the exterior membrane tension near the butterfly point, which are not determined by the leading chaos-related pole-skipping momentum alone. In this respect, the nonlinear drag expansion resolves additional near-horizon structure beyond the scale captured by the butterfly velocity.

The same reduced geometry also enters certain properties of stochastic heavy-quark dynamics. In an isotropic trailing-string background, fluctuations of the string worldsheet determine the transverse and longitudinal momentum-broadening coefficients. Denoting the magnitude of the drag force by \(f(v)=\lvert F(v)\rvert\), these coefficients satisfy
\begin{equation}
\kappa_\perp
=
2T_{\mathrm{ws}}\frac{f(v)}{v},
\qquad
\kappa_\parallel
=
2T_{\mathrm{ws}}\frac{\dd f(v)}{\dd v},
\label{eq:Langevincoefficients}
\end{equation}
where \(T_{\mathrm{ws}}\) is the Hawking temperature of the induced worldsheet horizon \cite{GiecoldIancuMueller2009,GiataganasSoltanpanahi2014}.

Although the individual broadening coefficients depend on the worldsheet temperature, this common factor cancels in their ratio. For the common-frame force, we obtain
\begin{equation}
\frac{\kappa_\parallel}{\kappa_\perp}
=
\frac{\dd\ln f}{\dd\ln v}
=
1+\frac{2v^2}{p\Delta(v^2)} .
\label{eq:Langevinratio}
\end{equation}
The longitudinal-to-transverse broadening ratio is therefore determined by the same causal stiffness \(\Delta(x)\) that governs the normalized drag force. This provides an additional example in which a quantity associated with worldsheet fluctuations can be expressed entirely in terms of the reduced geometric data reconstructed from the membrane.

There is, however, an important distinction between this ratio and the individual broadening coefficients. The absolute noise strengths retain their dependence on \(T_{\mathrm{ws}}\), which is sensitive to the radial metric component not captured by the geometric membrane. The correspondence therefore determines the anisotropy of momentum broadening in the common-frame setting, but not its overall scale.

This observation illustrates a more general limitation of the reconstruction. The geometric membrane and the stationary drag force depend on the causal-area relation \(a(Y)\), but do not determine the independent radial proper-distance element of the bulk metric. To see this, consider a smooth deformation of \(K(r)\) while keeping \(D(r)\), \(C(r)\), and the effective string coupling \(Z(r)\) fixed. Such a deformation leaves the worldsheet-horizon condition and the conserved momentum flux unchanged, and likewise does not affect the geometric membrane tension. Observables involving radial propagation or integration, however, generally remain sensitive to this deformation.

Examples include lightlike Wilson-loop integrals, generic quasinormal spectra, screening lengths, and the worldsheet Hawking temperature. The membrane--drag correspondence should therefore be understood as an exact relation between observables determined by the same algebraically reduced geometry, rather than as a reconstruction of all physical properties of the black-brane background. The distinction between causal-area data and radial-lapse information is also reflected in the scalar pole-skipping reconstruction, where the spectral tower contains geometric data that do not enter the stationary drag force.

A further subtlety appears in the special case \(p=1\), corresponding to a two-dimensional boundary theory. For the planar BTZ black brane, the causal and transverse-area functions are
\begin{equation}
a(z)=1-z^2,
\qquad
Y=z^{-2}.
\label{eq:BTZdata}
\end{equation}
The membrane relations then reduce to
\begin{equation}
q=1,
\qquad
M=1.
\label{eq:BTZmembrane}
\end{equation}
In this case, the squared membrane slope is independent of the radial position. The radial map therefore ceases to be invertible, and the differentiable Legendre reconstruction used for a generic membrane branch becomes singular.

Nevertheless, the trailing-string drag force remains well defined. The corresponding variational transformation can still be interpreted through its endpoint value, giving
\begin{equation}
H(x)
=
\frac{1}{1-x},
\qquad
F(v)
=
-\frac{\sigma v}{1-v^2}.
\label{eq:BTZdrag}
\end{equation}
The BTZ example is instructive because it separates the existence of a well-defined drag force from the invertibility of the membrane parametrization. The degeneracy arises because the conventional membrane tension no longer distinguishes different radial positions, rather than because the underlying trailing-string construction fails. This special behavior is consistent with the need for a generalized membrane description in two-dimensional conformal field theories.

It is also useful to comment on the structural assumptions underlying the exact tangent transformation. The central identity \(M'(q)=Y\) follows from the fact that the squared off-shell membrane density is affine in the squared membrane slope,
\begin{equation}
\mathscr{L}(r,q)
=
Y(r)\left[q-a(r)\right].
\label{eq:affinecriterion}
\end{equation}
This affine dependence allows the radial extremization to be interpreted as a Legendre transformation, with the derivative of the on-shell squared tension reconstructing the transverse-area variable. The subsequent relation to drag follows because the worldsheet-horizon condition selects the same reduced causal function.

A similar construction may remain possible in more general theories if the entropy functional reduces, in the appropriate scaling limit, to an affine expression of the form
\begin{equation}
\mathscr{L}(r,q)
=
A(r)q-B(r),
\label{eq:generalAffine}
\end{equation}
provided that the worldsheet horizon is determined by the ratio \(B/A\) and that the local effective string tension is a known function of \(A\). Under these conditions, radial extremization again produces a canonical relation that can support an analogous tangent transformation.

In generic higher-curvature theories, however, the holographic entropy functional may contain nonlinear dependence on the membrane slope, additional extrinsic-curvature contributions, or other geometric structures that cannot be absorbed into a single auxiliary function. The on-shell membrane tension need not then determine the transverse warp factor through a simple Legendre derivative. Although the membrane framework can remain applicable in the presence of finite-coupling corrections \cite{MezeiVirrueta2019}, the relation to stationary drag may require a generalized canonical construction adapted to the particular gravitational theory.

These possibilities suggest that the affine structure of the off-shell membrane density, rather than the detailed form of the black-brane metric, is the essential ingredient behind the simple membrane--drag transformation. Identifying the corresponding canonical structure in more general holographic theories would provide a natural way to examine how much of the geometric correspondence persists beyond the two-derivative Einstein-gravity setting.

\subsection*{Conclusions}
\label{sec:conclusions}

We have established an exact correspondence between the geometric entanglement membrane and the stationary drag force of a heavy probe in an isotropic holographic thermal state. The correspondence is governed by the causal-area relation $a(Y)$, which survives the algebraic elimination of the radial coordinate in both variational problems. For the membrane, this elimination produces a Legendre structure with $M'(q)=Y$; for the trailing string, the worldsheet horizon selects the point at which $a=v^2$. Combining these relations yields an invertible tangent transformation between the exterior geometric membrane and the normalized common-frame drag curve. The construction identifies a shared sector of bulk geometry without equating the two observables or their respective velocity variables.

The small-velocity expansion makes this shared information accessible order by order. The coefficient of $v^3$ is fixed by the butterfly velocity, whereas the coefficient of $v^5$ determines the membrane curvature at the butterfly point. Higher corrections successively resolve higher derivatives of the tension, giving an all-order local reconstruction of the membrane near the horizon. Scalar pole-skipping supplies an independent route to the same data. By combining the algebraic reconstruction of the near-horizon metric with the drag expansion, we express the first nonlinear coefficients explicitly in terms of skipped momenta and extend the relation through the entire scalar tower. The resulting correspondence connects nonlinear defect response to spectral information in an ambient scalar two-point function. It also separates the causal-area data entering mean drag from the additional radial-lapse data encoded in scalar propagation.

The null energy condition constrains this common geometry through a nonnegative causal load. Its integral measures the reduction of the butterfly velocity from the AdS--Schwarzschild value, while its horizon value and successive derivatives are resolved by nonlinear drag coefficients. The same restriction gives the minimum normalized drag bound and the associated membrane and information-velocity inequalities. These comparisons require a fixed horizon and probe normalization; they are not statements that Schwarzschild minimizes the absolute drag at an arbitrary fixed temperature. The charged examples illustrate how matter modifies the complete response, while their extremal limits demonstrate that the exact transformation can remain meaningful even when the ordinary Taylor expansion is replaced by a nonanalytic one.

The relation to information recovery extends the physical interpretation of the coefficient hierarchy. Near maximal initial entanglement, the quintic drag coefficient determines the first correction to the strip information velocity, and higher coefficients determine its successive endpoint derivatives. Knowledge of the complete information-velocity function goes further: it reconstructs the physical interior membrane and the causal-area relation from the horizon down to the entanglement bottleneck. This interior reconstruction complements the exterior reconstruction from physical drag, with the two descriptions meeting at the butterfly point. It requires a family of recovery protocols governed by the same final-state membrane and does not determine the radial lapse or regions deeper than the bottleneck. Without additional analyticity assumptions, complete exterior data or an infinite horizon Taylor jet do not by themselves fix an arbitrary smooth interior continuation.

The missing exterior radial information can instead be supplied by stochastic heavy-quark dynamics. The absolute transverse momentum broadening determines the worldsheet Hawking temperature once the mean force is known, and this temperature fixes the radial proper-distance function. A complete velocity scan of mean drag and one absolute broadening coefficient therefore reconstructs the full exterior metric, up to its overall scale, within the specified classical common-frame probe theory. The longitudinal broadening is then predicted by the same drag curve and provides a consistency check. For a radially dressed string, the corresponding inversion reconstructs the effective string-frame metric; separating it into an Einstein-frame geometry and an independent coupling function requires additional information.

Several extensions are suggested by this structure. Anisotropic or time-dependent backgrounds would test whether the one-variable tangent construction can be replaced by an appropriate multivariable relation. Higher-curvature theories require identifying which features of the entropy functional preserve the affine dependence on the squared membrane slope that underlies the present Legendre structure. Higher-gradient membrane terms and finite-frequency probe fluctuations may provide complementary ways of resolving geometric information lost in the leading scaling limit. These extensions require their own dynamical analysis, rather than a direct extrapolation of the common-frame Einstein formulas.

The central conclusion is that entanglement dynamics, nonlinear heavy-probe response, information recovery, and scalar pole-skipping encode overlapping but distinguishable parts of the bulk geometry. The membrane provides a common set of variables in which these overlaps can be calculated explicitly, while probe dressing and radial propagation identify the information that remains independent. This turns the membrane from an effective description of entropy growth into a useful organizing principle for reconstructing geometry from complementary boundary observables\footnote{The author used AI tools during work on this manuscript. Author
reviewed and edited the content as needed and take full responsibility for the content of the publication.}.

\newpage
\bibliographystyle{JHEP}
\bibliography{biblio}

@article{RyuTakayanagi2006,
    author = "Ryu, Shinsei and Takayanagi, Tadashi",
    title = "{Holographic derivation of entanglement entropy from AdS/CFT}",
    eprint = "hep-th/0603001",
    archivePrefix = "arXiv",
    primaryClass = "hep-th",
    doi = "10.1103/PhysRevLett.96.181602",
    journal = "Phys. Rev. Lett.",
    volume = "96",
    pages = "181602",
    year = "2006"
}

@article{HubenyRangamaniTakayanagi2007,
    author = "Hubeny, Veronika E. and Rangamani, Mukund and Takayanagi, Tadashi",
    title = "{A Covariant holographic entanglement entropy proposal}",
    eprint = "0705.0016",
    archivePrefix = "arXiv",
    primaryClass = "hep-th",
    doi = "10.1088/1126-6708/2007/07/062",
    journal = "JHEP",
    volume = "07",
    pages = "062",
    year = "2007"
}

@article{Mezei2018,
    author = "Mezei, M{\'a}rk",
    title = "{Membrane theory of entanglement dynamics from holography}",
    eprint = "1803.10244",
    archivePrefix = "arXiv",
    primaryClass = "hep-th",
    doi = "10.1103/PhysRevD.98.106025",
    journal = "Phys. Rev. D",
    volume = "98",
    number = "10",
    pages = "106025",
    year = "2018"
}

@article{HerzogEtAl2006,
    author = "Herzog, Christopher P. and Karch, Andreas and Kovtun, Pavel and Kozcaz, Can and Yaffe, Laurence G.",
    title = "{Energy loss of a heavy quark moving through N=4 supersymmetric Yang-Mills plasma}",
    eprint = "hep-th/0605158",
    archivePrefix = "arXiv",
    primaryClass = "hep-th",
    doi = "10.1088/1126-6708/2006/07/013",
    journal = "JHEP",
    volume = "07",
    pages = "013",
    year = "2006"
}

@article{Arefeva:2014QGP,
    author  = "Aref'eva, I. Ya.",
    title   = "{Holographic approach to quark-gluon plasma in heavy ion collisions}",
    journal = "Phys. Usp.",
    volume  = "57",
    pages   = "527--555",
    year    = "2014",
    doi     = "10.3367/UFNe.0184.201406a.0569"
}

@article{Gubser2006,
    author = "Gubser, Steven S.",
    title = "{Drag force in AdS/CFT}",
    eprint = "hep-th/0605182",
    archivePrefix = "arXiv",
    primaryClass = "hep-th",
    doi = "10.1103/PhysRevD.74.126005",
    journal = "Phys. Rev. D",
    volume = "74",
    pages = "126005",
    year = "2006"
}

@article{Ageev2021,
    author = "Ageev, Dmitry S.",
    title = "{Chaotic nature of holographic QCD}",
    eprint = "2105.04589",
    archivePrefix = "arXiv",
    primaryClass = "hep-th",
    doi = "10.1103/PhysRevD.104.126013",
    journal = "Phys. Rev. D",
    volume = "104",
    number = "12",
    pages = "126013",
    year = "2021"
}

@article{Mezei2017,
    author = "Mezei, M{\'a}rk",
    title = "{On entanglement spreading from holography}",
    eprint = "1612.00082",
    archivePrefix = "arXiv",
    primaryClass = "hep-th",
    doi = "10.1007/JHEP05(2017)064",
    journal = "JHEP",
    volume = "05",
    pages = "064",
    year = "2017"
}

@article{GursoyEtAl2009,
    author = "G{\"u}rsoy, Umut and Kiritsis, Elias and Michalogiorgakis, Georgios and Nitti, Francesco",
    title = "{Thermal transport and drag force in improved holographic QCD}",
    eprint = "0906.1890",
    archivePrefix = "arXiv",
    primaryClass = "hep-ph",
    doi = "10.1088/1126-6708/2009/12/056",
    journal = "JHEP",
    volume = "12",
    pages = "056",
    year = "2009"
}

@article{BlakeDavisonVegh2020,
    author = "Blake, Mike and Davison, Richard A. and Vegh, David",
    title = "{Horizon constraints on holographic Green's functions}",
    eprint = "1904.12883",
    archivePrefix = "arXiv",
    primaryClass = "hep-th",
    doi = "10.1007/JHEP01(2020)077",
    journal = "JHEP",
    volume = "01",
    pages = "077",
    year = "2020"
}

@article{BlakeDavisonGrozdanovLiu2018,
    author = "Blake, Mike and Davison, Richard A. and Grozdanov, Sa{\v{s}}o and Liu, Hong",
    title = "{Many-body chaos and energy dynamics in holography}",
    eprint = "1809.01169",
    archivePrefix = "arXiv",
    primaryClass = "hep-th",
    doi = "10.1007/JHEP10(2018)035",
    journal = "JHEP",
    volume = "10",
    pages = "035",
    year = "2018"
}

@article{LuRanWu2026,
    author = "Lu, Zhenkang and Ran, Cheng and Wu, Shao-Feng",
    title = "{Algebraic structure underlying pole-skipping points}",
    eprint = "2507.13306",
    archivePrefix = "arXiv",
    primaryClass = "hep-th",
    doi = "10.1103/lgkz-gyl1",
    journal = "Phys. Rev. D",
    volume = "113",
    number = "4",
    pages = "046008",
    year = "2026"
}

@article{WangWang2022,
    author = "Wang, Diandian and Wang, Zi-Yue",
    title = "{Pole Skipping in Holographic Theories with Bosonic Fields}",
    eprint = "2208.01047",
    archivePrefix = "arXiv",
    primaryClass = "hep-th",
    doi = "10.1103/PhysRevLett.129.231603",
    journal = "Phys. Rev. Lett.",
    volume = "129",
    number = "23",
    pages = "231603",
    year = "2022"
}

@article{AhnEtAl2021,
    author = "Ahn, Yongjun and Jahnke, Viktor and Jeong, Hyun-Sik and Lee, Kyung-Sun and Nishida, Mitsuhiro and Kim, Keun-Young",
    title = "{Classifying pole-skipping points}",
    eprint = "2010.16166",
    archivePrefix = "arXiv",
    primaryClass = "hep-th",
    doi = "10.1007/JHEP03(2021)175",
    journal = "JHEP",
    volume = "03",
    pages = "175",
    year = "2021"
}

@article{GiecoldIancuMueller2009,
    author = "Giecold, Gregory C. and Iancu, Edmond and Mueller, Alfred H.",
    title = "{Stochastic trailing string and Langevin dynamics from AdS/CFT}",
    eprint = "0903.1840",
    archivePrefix = "arXiv",
    primaryClass = "hep-th",
    doi = "10.1088/1126-6708/2009/07/033",
    journal = "JHEP",
    volume = "07",
    pages = "033",
    year = "2009"
}

@article{GiataganasSoltanpanahi2014,
    author = "Giataganas, Dimitrios and Soltanpanahi, Hesam",
    title = "{Universal properties of the Langevin diffusion coefficients}",
    eprint = "1310.6725",
    archivePrefix = "arXiv",
    primaryClass = "hep-th",
    doi = "10.1103/PhysRevD.89.026011",
    journal = "Phys. Rev. D",
    volume = "89",
    number = "2",
    pages = "026011",
    year = "2014"
}

@article{MezeiVirrueta2019,
    author = "Mezei, M{\'a}rk and Virrueta, Julio",
    title = "{Exploring the Membrane Theory of Entanglement Dynamics}",
    eprint = "1912.11024",
    archivePrefix = "arXiv",
    primaryClass = "hep-th",
    doi = "10.1007/JHEP02(2020)013",
    journal = "JHEP",
    volume = "02",
    pages = "013",
    year = "2020"
}

@article{CouchEtAl2020,
    author = "Couch, Josiah and Eccles, Stefan and Nguyen, Phuc and Swingle, Brian and Xu, Shenglong",
    title = "{Speed of quantum information spreading in chaotic systems}",
    eprint = "1908.06993",
    archivePrefix = "arXiv",
    primaryClass = "cond-mat.stat-mech",
    doi = "10.1103/PhysRevB.102.045114",
    journal = "Phys. Rev. B",
    volume = "102",
    number = "4",
    pages = "045114",
    year = "2020"
}

@article{EcclesEtAl2021,
    author = "Eccles, Stefan and Fischler, Willy and Guglielmo, Tyler and Pedraza, Juan F. and Racz, Sarah",
    title = "{Speeding up the spread of quantum information in chaotic systems}",
    eprint = "2108.12688",
    archivePrefix = "arXiv",
    primaryClass = "hep-th",
    doi = "10.1007/JHEP12(2021)019",
    journal = "JHEP",
    volume = "12",
    pages = "019",
    year = "2021"
}

@article{JiangMezeiVirrueta2025,
    author = "Jiang, Hanzhi and Mezei, M{\'a}rk and Virrueta, Julio",
    title = "{The entanglement membrane in 2d CFT: reflected entropy, RG flow, and information velocity}",
    eprint = "2411.16542",
    archivePrefix = "arXiv",
    primaryClass = "hep-th",
    doi = "10.1007/JHEP06(2025)114",
    journal = "JHEP",
    volume = "06",
    pages = "114",
    year = "2025"
}

@article{CrossleyGloriosoLiu2017,
    author = "Crossley, Michael and Glorioso, Paolo and Liu, Hong",
    title = "{Effective field theory of dissipative fluids}",
    eprint = "1511.03646",
    archivePrefix = "arXiv",
    primaryClass = "hep-th",
    doi = "10.1007/JHEP09(2017)095",
    journal = "JHEP",
    volume = "09",
    pages = "095",
    year = "2017"
}

@article{VanRaamsdonk2010,
    author = "Van Raamsdonk, Mark",
    title = "{Building up spacetime with quantum entanglement}",
    eprint = "1005.3035",
    archivePrefix = "arXiv",
    primaryClass = "hep-th",
    doi = "10.1007/s10714-010-1034-0",
    journal = "Gen. Rel. Grav.",
    volume = "42",
    pages = "2323--2329",
    year = "2010"
}

@article{AlmheiriDongHarlow2015,
    author = "Almheiri, Ahmed and Dong, Xi and Harlow, Daniel",
    title = "{Bulk Locality and Quantum Error Correction in AdS/CFT}",
    eprint = "1411.7041",
    archivePrefix = "arXiv",
    primaryClass = "hep-th",
    doi = "10.1007/JHEP04(2015)163",
    journal = "JHEP",
    volume = "04",
    pages = "163",
    year = "2015"
}

@article{DongHarlowWall2016,
    author = "Dong, Xi and Harlow, Daniel and Wall, Aron C.",
    title = "{Reconstruction of Bulk Operators within the Entanglement Wedge in Gauge-Gravity Duality}",
    eprint = "1601.05416",
    archivePrefix = "arXiv",
    primaryClass = "hep-th",
    doi = "10.1103/PhysRevLett.117.021601",
    journal = "Phys. Rev. Lett.",
    volume = "117",
    number = "2",
    pages = "021601",
    year = "2016"
}

@article{HartmanMaldacena2013,
    author = "Hartman, Thomas and Maldacena, Juan",
    title = "{Time Evolution of Entanglement Entropy from Black Hole Interiors}",
    eprint = "1303.1080",
    archivePrefix = "arXiv",
    primaryClass = "hep-th",
    doi = "10.1007/JHEP05(2013)014",
    journal = "JHEP",
    volume = "05",
    pages = "014",
    year = "2013"
}

@article{LiuSuh2014Tsunami,
    author = "Liu, Hong and Suh, S. Josephine",
    title = "{Entanglement Tsunami: Universal Scaling in Holographic Thermalization}",
    eprint = "1305.7244",
    archivePrefix = "arXiv",
    primaryClass = "hep-th",
    doi = "10.1103/PhysRevLett.112.011601",
    journal = "Phys. Rev. Lett.",
    volume = "112",
    pages = "011601",
    year = "2014"
}

@article{LiuSuh2014Growth,
    author = "Liu, Hong and Suh, S. Josephine",
    title = "{Entanglement growth during thermalization in holographic systems}",
    eprint = "1311.1200",
    archivePrefix = "arXiv",
    primaryClass = "hep-th",
    doi = "10.1103/PhysRevD.89.066012",
    journal = "Phys. Rev. D",
    volume = "89",
    number = "6",
    pages = "066012",
    year = "2014"
}

@article{NahumEtAl2017,
    author = "Nahum, Adam and Ruhman, Jonathan and Vijay, Sagar and Haah, Jeongwan",
    title = "{Quantum Entanglement Growth Under Random Unitary Dynamics}",
    eprint = "1608.06950",
    archivePrefix = "arXiv",
    primaryClass = "cond-mat.stat-mech",
    doi = "10.1103/PhysRevX.7.031016",
    journal = "Phys. Rev. X",
    volume = "7",
    number = "3",
    pages = "031016",
    year = "2017"
}

@article{JonayHuseNahum2018,
    author = "Jonay, Cheryne and Huse, David A. and Nahum, Adam",
    title = "{Coarse-grained dynamics of operator and state entanglement}",
    eprint = "1803.00089",
    archivePrefix = "arXiv",
    primaryClass = "cond-mat.stat-mech",
    month = "2",
    year = "2018"
}

@article{Arefeva:2020QFT,
    author  = "Aref'eva, I.",
    title   = "{Holography for Nonperturbative Study of QFT}",
    journal = "Phys. Part. Nucl.",
    volume  = "51",
    number  = "4",
    pages   = "489--496",
    year    = "2020",
    doi     = "10.1134/S1063779620040097"
}

@article{Sin:2006pv,
    author        = "Sin, Sang-Jin and Zahed, Ismail",
    title         = "{Ampere's Law and Energy Loss in AdS/CFT Duality}",
    eprint        = "hep-ph/0606049",
    archivePrefix = "arXiv",
    journal       = "Phys. Lett. B",
    volume        = "648",
    pages         = "318--322",
    year          = "2007",
    doi           = "10.1016/j.physletb.2007.01.074"
}

@article{Andreev:2017Drag,
    author        = "Andreev, Oleg",
    title         = "{Drag Force on Heavy Quarks and Spatial String Tension}",
    eprint        = "1707.05045",
    archivePrefix = "arXiv",
    primaryClass  = "hep-ph",
    journal       = "Mod. Phys. Lett. A",
    volume        = "33",
    number        = "06",
    pages         = "1850041",
    year          = "2018",
    doi           = "10.1142/S0217732318500414"
}

\end{document}